\documentclass[%
reprint,
superscriptaddress,
nofootinbib,
amsmath,
amssymb,
aps,
floatfix,
showkeys,
]{revtex4-2}

\usepackage[colorlinks,citecolor=blue,linkcolor=blue,anchorcolor=blue,filecolor=blue,
urlcolor=blue]{hyperref}

\usepackage{graphicx,color,xcolor}
\usepackage{amsfonts,amsmath}
\usepackage{amssymb,ulem}
\usepackage{dcolumn}
\usepackage{bm,lipsum}
\usepackage[utf8]{inputenc}
\usepackage{subeqnarray}
\usepackage{array}
\usepackage{epsfig,hyperref}
\usepackage{morefloats}
\usepackage{relsize}
\usepackage{url}
\usepackage{orcidlink}

\usepackage{comment}

\begin{document}

\preprint{ }

\title{Global and crustal properties of neutron stars}

\author{Luiz L. Lopes}

\affiliation{ Centro Federal de Educac\~{a}o Tecnol\'{o}gica de Minas Gerais Campus VIII; CEP 37.022-560, Varginha - MG - Brazil\\}
% \mbox{$^2$Depto de F\'{\i}sica - CFM - Universidade Federal de Santa Catarina  Florian\'opolis - SC - CP. 476 - CEP 88.040 - 900 - Brazil }}

\author{Guilherme \surname{Grams}}
\affiliation{Institut f\"ur Physik und Astronomie, Universit\"at Potsdam, Haus 28, Karl-Liebknecht-Str. 24/25, 14476, Potsdam, Germany}
\date{\today}

\begin{abstract}
We investigate how the treatment of the neutron star crust affects both the global properties of the star and the properties of the crust itself. 
Three approaches are compared: a direct connection between a fixed crust and the core equation of state (EOS), a smooth crossover between them,
and a unified EOS in which the crust is computed within a compressible liquid-drop model with the same 
%isoscalar and isovector physical 
nuclear empirical parameters 
%built with 
of the relativistic mean-field model used for the core. 
For each approach, we consider four values of the symmetry energy slope, $44 \le L \le 92$ MeV. For a $1.4\,M_\odot$ star, some global properties are weakly to moderately affected: the radius and the moment of inertia agree within  1\% among the three approaches. On the other hand,  the tidal deformability obtained with the direct connection is up to 21\% lower than the unified one for $L = 92$ MeV. 
A much larger contrast comes from the crustal properties.
Relative to the unified EOS, the direct connection yields a crust that is 15--36\% thinner, 6--12 times less massive, and that carries a fraction of the total moment of inertia 5--8 times smaller (0.3--1.1\% instead of 2.8--6.2\%). The differences between crust treatments are thus comparable to or larger than those arising from variations in $L$ for crustal properties, while they remain subleading for global ones.
\end{abstract}

\maketitle

\section{Introduction}

Neutron stars (NS) are the densest stars in the Universe, which makes them the perfect laboratory for connecting nuclear physics and astrophysics. The NS structure can be divided into layers, and many nuclear-physics inputs are essential. The surface is believed to be covered with $^{56}$Fe and neighboring elements. The solid outer crust is composed of highly exotic nuclei (embedded in a free electron gas) that progressively become more neutron-rich with increasing depth until unbound neutrons appear. This marks the transition to the inner crust, where neutron-rich nuclei coexist with a neutron (super)fluid. Therefore, understanding the properties of nuclear matter and finite nuclei is essential for the crust of NS. 
In the core, nucleons can no longer form clusterized matter, and neutrons, protons, and leptons form a homogeneous liquid in beta equilibrium~\cite {Haensel2007}. 

The crust is thus the region of the NS where the connection between terrestrial nuclear physics and astrophysics is most direct. 
Matter is still organized into nuclei, and the outermost layers are made of isotopes that can be produced and studied in the laboratory, so that their composition and EOS are determined almost model independently by measured atomic masses~\cite{Chamel2008, Haensel2007, Blaschke2018}. Mass
measurements of exotic neutron-rich nuclei have progressively extended this experimental knowledge deeper into the outer crust~\cite{Wolf2013}. 
Beyond the neutron drip, the neutron fluid stabilizes clusters that are much heavier and more neutron-rich than any nucleus produced in terrestrial laboratories, and the description must rely on nuclear models~\cite{Chamel2008, Blaschke2018}. 
Describing this sequence consistently therefore requires a model that is anchored to experimental nuclear masses at low density and that can be extrapolated, together with the properties of homogeneous matter, toward the core.

%[some general reviews and study of crust properties [Chamel, Newton, Grams, Göttling, Thomas] + studies of crust-core matching [Fortin, Constanca]]
A comprehensive review of the physics of NS crusts can be found in Ref.~\cite{Chamel2008}. Over the last years, considerable effort has
been devoted to quantifying how uncertainties in the nuclear physics input
propagate into crustal properties. 
Compressible liquid-drop models informed by nuclear-matter empirical parameters and experimental masses have been
used to correlate the crust composition, the neutron-drip and crust-core
transition points, and the crust thickness with the properties of nuclear
matter~\cite{Grams2022PRC, Grams2022EPJA}, to perform Bayesian analyses of
the crust-core transition~\cite{Carreau2019EPJA}, and to establish general
predictions for the crustal moment of inertia~\cite{Carreau2019PRC}.
Bayesian frameworks have also been employed to construct prior distributions of crust models, including the nuclear pasta phases~\cite{Balliet2021}, and inner-crust EOSs consistent with chiral effective field theory, with quantified theoretical uncertainties, have recently become available~\cite{Gottling2026}. 
In parallel, it has been shown that non-unified EOSs, in which a standard crust is attached to an independently derived core EOS, introduce spurious uncertainties in the stellar radius~\cite{Fortin2016}, that the results depend on the matching procedure itself~\cite{Ferreira2020Universe}, and that the treatment of the crust can even affect the empirical relations used to interpret observations~\cite{Ferreira2020PRD}. 
Matching procedures nevertheless remain common in practice, particularly with relativistic mean-field models of the core. It is therefore worth assessing, within a single nuclear model, how the choice of crust treatment compares with the uncertainty stemming from the density dependence of the symmetry energy, both for the crust itself and for the star as a whole.

These uncertainties will soon become observationally relevant.
Next-generation gravitational-wave observatories, such as the Einstein Telescope~\cite{ET2026} and Cosmic Explorer~\cite{Reitze2019Cosmic}, will considerably improve these measurements. 
For a single binary NS event, the radius of a NS is expected to be measured with uncertainties of about 200, 100, and 50~m for networks with one, two, and three such detectors, respectively~\cite{Huxford2024}. 
This precision is comparable to the radius differences between unified and
non-unified EOSs~\cite{Fortin2016, Ferreira2020Universe}, which also affect the tidal deformability~\cite{Pang2026PRD}.
In the electromagnetic domain, the enhanced X-ray Timing and Polarimetry mission (eXTP), planned for launch in 2030, will extend the pulse-profile modelling pioneered by NICER~\cite{NICER2012,Salmi_2024, Choudhury2024}, to a larger and more diverse population of NSs~\cite{Li2025eXTP}.

The consequences of the crust treatment, however, go beyond the stellar radius. 
Several observed phenomena are governed by the crust itself rather than by the global structure of the star. Pulsar glitches are commonly attributed to angular momentum stored in the superfluid neutrons of the
inner crust, which requires the crust to carry a minimum fraction of the total moment of inertia~\cite{Link1999}, a requirement that becomes
considerably more demanding once the entrainment of free neutrons by the crustal lattice is taken into account~\cite{Andersson2012, Chamel2013PRL}.
The quasi-periodic oscillations detected in magnetar giant flares have been interpreted as torsional modes of the crust~\cite{Israel2005, Strohmayer2005, Steiner2009}, and the thermal relaxation of accreting
NS after an outburst probes the thickness and composition of
the crust~\cite{Brown2009}. 
All of these depend on the crustal mass, thickness, and moment of inertia, which are set mainly by the density and
pressure at the crust-core transition. 
These quantities may therefore be far more sensitive to how the crust is modelled than global observables such as the radius or the total moment of inertia, and it is natural to ask whether the simplifications adopted in non-unified EOSs, harmless for the latter, remain acceptable for the former.

We find that the crust treatment is a subleading source of uncertainty for the global properties of the star, but a dominant one for its crustal properties, to the point of determining how the latter depend on $L$.

This paper is organized as follows: in Sec.~\ref{sec:core} we present the formalism and parametrization for the core EOS, while in Sec.~\ref{sec:crust} we present the formalism for the three approaches for the crust. In Sec.~\ref{sec:micro} we present the microphysics results, and in Sec.~\ref{sec:macro} the macroscopic properties, divided into two parts: global and crustal.

%%%%%%%%%%%%%%%%%%%%%%%%%%%%%%%%%%%%%%%%%%%%%%
\section{The core EOS: QHD}
\label{sec:core}

In the NS core, matter is homogeneous, which simplifies calculations once we disregard surface effects and Coulomb-field contributions. To model the interaction between nucleons, we use a non-linear version of Quantum Hadrodynamics (QHD) in the mean-field approximation (MFA)~\cite{Serot_1992}. Furthermore, as NS are charge-neutral, chemically stable objects, leptons are added as a relativistic free Fermi gas. The total Lagrangian density in MFA reads~\cite{Miyatsu2013}:
\begin{eqnarray}
\mathcal{L} = \sum_B\bar{\psi}_B[\gamma^0(i\partial_0 -g_{N\omega}\omega_0 - \frac{1}{2}g_{N\rho}{\tau_3} {\rho}_0) \nonumber \\ -i\gamma^j \partial_j - M_B^*]\psi_B -\frac{1}{2}m_s^2\sigma_0^2    + \frac{1}{2}m_\omega^2\omega_0^2
+ \frac{1}{2}m_\rho^2{\rho}_0^2 \nonumber \\ - \frac{\kappa M_B (g_{Ns}\sigma_0)^3}{3} - \frac{\lambda (g_{NS}\sigma_0)^4}{4}\nonumber  + \Lambda_{\omega\rho}(g_{N\rho}^2g_{N\omega}^2{\rho_0^2}\omega_0^2) \\  +
\sum_l \bar{\psi}_l[(i\gamma^\mu\partial_\mu - m_l)\psi_l \label{totalL},
\end{eqnarray}
where the sum in $B$ runs over protons and neutrons, $\psi_B$ is the Dirac field of the nucleons, the g's are the Yukawa couplings to simulate the strong force, $M_B^{*}$ is the effective nucleon mass, $M_B^{*} = M_B - g_{N\sigma}\sigma_0$, with $M_B = 939$ MeV, for both, protons and neutrons; $\sigma_0$, $\omega_0$, and $\rho_0$ are the mesonic fields in MFA, and $m_i's$ are their respective masses, and $\psi_l$ and $m_l$ are the Dirac field and the masses of the two lightest charged leptons, electrons and muons.

A complete and detailed derivation of the EOS for the QHD in MFA can be seen in Ref.~\cite{LopesUNIVERSE2025}. Now, any realistic EOS must be able to fulfill the constraints of at least five well-known parameters at the saturation point: the saturation point itself, $n_0$, the binding energy per nucleon, $B/A$, the incompressibility of the nuclear matter, $K_0$, the nucleon effective mass, $M^*/M$, and the symmetry energy $S_0$. To fulfill this task, we employ here the L3$\omega\rho$ model, introduced in Ref.~\cite{Lopes2022CTP}. The parametrization and the nuclear constraints at the saturation density are presented in Tab.~\ref{TL1}, with the constraints taken from two extensive review articles, Refs.~\cite{Dutra2014,Micaela2017}.

\begin{table}[h]
\begin{center}
\begin{tabular}{cc|cccc}
\toprule
  & Parameters & &  Constraints  & This model  \\
\toprule
 $\left(g_{N\sigma}/{m_s}\right)^2$ & $12.108 \, \mathrm{fm}^2$ &$n_0 (\mathrm{fm}^{-3})$ & 0.148 - 0.170 & 0.156 \\
% \hline
$\left(g_{N\omega}/{m_v}\right)^2$  & $7.132 \, \mathrm{fm}^2$ & $M^{*}/M$ & 0.6 - 0.8 & 0.69  \\
%  \hline
  $\kappa$ & 0.004138 & $K_0 \mathrm{(MeV)}$ & 220 - 260             &  256  \\
% \hline
$\lambda$ &  -0.00390  & $S_0 \mathrm{(MeV)}$  & 28.6 - 34.4 &  31.7  \\
%\hline
- &  - & -$B/A \mathrm{(MeV)}$  & 15.8 - 16.5  & 16.2 \\
\toprule
\end{tabular}
\caption{Model parameters used in this study and their predictions for symmetric nuclear matter at saturation density. The phenomenological constraints were taken from Refs.~\cite{Dutra2014, Micaela2017}.}
\label{TL1}
\end{center}
\end{table}

A sixth physical quantity, the symmetry energy slope, $L$, has gained considerable attention nowadays, once it is linked to measurable features, like the neutron skin~\cite{IUFSU}, and the radius of NS~\cite{Rafa2011,lopescesar}.

In order to vary the slope while keeping all the physical quantities presented in Tab.~\ref{TL1} constant, we change only the parameters $(g_{N\rho}/m_\rho)^2$, and $\Lambda_{\omega\rho}$. The values used for these constants, and the predicted values for the slope, are presented in Tab.~\ref{TL2}. Values larger than L=$L = 92$ MeV will not be used here, as they were ruled out in Refs.~\cite{lopescesar,lopes2024PRC} for the present parametrization, although some experimental results suggest values as high as 143 MeV~\cite{PREX2}.

%------------------------------------------------------
\begin{center}
\begin{table}[bt]
\begin{tabular}{ccccc}
\toprule
  $L$ (MeV) & $(g_\rho/m_\rho)^2$ ($\mathrm{fm}^2$)  ($\mathrm{fm}^2$)  & $\Lambda_{\omega\rho}$  \\
\toprule
 44 & 8.40 &  0.0515 \\
 %\hline
 60 & 6.16    & 0.0344 \\
%  \hline
 76& 4.90 &  0.0171        \\
% \hline
92 &  4.06   & 0  \\
%\hline
\toprule
\end{tabular}
\caption{Model parameters selected to set the symmetry energy at $S_0$ = 31.7 MeV}
\label{TL2}
\end{table}
\end{center}

%\section{The crust EOSs: The BBP; the CLDM + cuter; and the BBP + smooth crossover}
\section{The crust EOS}
\label{sec:crust}

The NS crust can be divided into two parts: the outer crust and the inner crust. 
The outer crust begins at the stellar surface, beneath a thin atmosphere of light elements whose thickness ranges from a few millimeters to some tens of centimeters depending on the surface temperature~\cite{Haensel2007}, and which does not contribute to the EOS.
At zero pressure, cold catalyzed matter consists of a body-centered cubic lattice of $^{56}$Fe at a density $\rho \approx  7.8$~g~cm$^{-3}$, corresponding to a baryon density $n \simeq 5\times10^{-15}$~fm$^{-3}$~\cite{Haensel2007}.
For $\rho \gtrsim 10^{4}$~g~cm$^{-3}$, atoms are fully ionized by pressure, and matter consists of nuclei embedded in a degenerate electron gas, which
becomes relativistic for $\rho \gtrsim 10^{6}$~g~cm$^{-3}$ and provides most of the pressure. 
As density increases, the electron Fermi energy rises, making electron captures energetically favorable, and nuclei become progressively more neutron-rich. The $^{56}$Fe layer extends up to $\rho \approx 8\times10^{6}$~g~cm$^{-3}$, and is followed by layers of Ni, Kr, Se, Ge, and Zn isotopes whose masses are experimentally known up to $\rho \sim 6\times10^{10}$~g~cm$^{-3}$; deeper layers must be described by nuclear physics models~\cite{Blaschke2018, Chamel2008}.

Assuming cold catalyzed matter, each layer of the outer crust forms a perfect crystal of a single nuclear species, with $N$ neutrons and $Z$ protons, embedded in the electron gas. The equilibrium composition at a given baryon density is obtained by minimizing the energy density (or the free energy density at finite temperature) with respect to the cluster variables, subject to baryon-number conservation and charge neutrality~\cite{Chamel2008, BurgioFantina2018}. 
In the inner crust, the same minimization also includes the density of the unbound neutron fluid
(Sec.~\ref{sec:crust}A). 
In the outer crust, the composition can equivalently be obtained by minimizing the Gibbs free energy per nucleon at fixed pressure, which is the procedure adopted in the seminal BPS EOS~\cite{BPS, Haensel2007}.
A detailed discussion of the transitions between successive nuclear species in the outer crust can be found in Ref.~\cite{Lopes2021EPL}.

In this work, we describe the outer crust with the BPS EOS in the direct-connection and smooth-crossover approaches. In the unified approach, we use the BPS EOS only below $10^{-4}$~fm$^{-3}$, and compute the deeper layers of the outer crust within the same model as the inner crust (Sec.~\ref{sec:crust}C).

As matter becomes increasingly neutron-rich, a point is reached at which it
becomes energetically favourable for some neutrons to escape from nuclei.
This point is known as neutron drip and marks the beginning of the inner
crust. Previous studies suggest 2.4 fm$^{-3}$ $< n_{\rm drip} <$ 3.6 fm$^{-3}$~\cite{Chamel2008}. Here, we use three approaches to model the deep layers of the outer crust, the NS inner crust, and its connection to the core.

%Microscopic nuclear mass models place it at
%$\rho_{\rm drip} \approx (4.2$--$4.5)\times10^{11}$~g~cm$^{-3}$, i.e.,
%$n_{\rm drip} \approx 2.6\times10^{-4}$~fm$^{-3}$~\cite{Blaschke2018}.
%Here, we use three different approaches to this phase.

\subsection{BBP+direct connection}

In this first approach, we use the Baym-Bethe-Pethick (BBP) EOS~\cite{BBP} for the whole inner crust up to a density of 0.01 fm$^{-3}$, and then direct connected with the core QHD EOS densities larger than that. This is the approach suggested in Glendenning's classical book, Ref.~\cite{Glenbook} and it is still used nowadays~\cite{Celi2024PRD,Lopes2024ApJ}.

The BBP EOS is described within the
compressible liquid-drop model (CLDM) using charge-neutral
Wigner--Seitz (WS) cells. In the spherical approximation, each cell
contains a nuclear cluster of radius $R_N$, baryon density $n_i$,
and proton fraction $x_i$, surrounded by a fluid of unbound neutrons
with density $n_f$ and an approximately uniform electron
background. Defining $u=(R_N/R_{\rm WS})^3$ as the fraction of
the cell volume occupied by the cluster, baryon conservation and
charge neutrality give
\begin{equation}
\bar n=u n_i+(1-u)n_f,
\qquad
n_e=u x_i n_i,
\end{equation}
where $\bar n$ is the mean baryon density. The model is termed
``compressible'' because the density inside clusters, $n_i$, is determined by energy
minimization rather than fixed at the saturation density.

At zero temperature, the total energy per unit volume of the
WS cell, including rest-mass contributions, is~\cite{BBP,Newton2013APJSS}
\begin{equation}
\epsilon_{\rm cell}
=u\,\epsilon_{\rm hom}(n_i,x_i)
+(1-u)\,\epsilon_{\rm hom}(n_f,0)
+\epsilon_e(n_e)
+\epsilon_{\rm surf}
+\epsilon_{\rm C}.
\end{equation}
The first term describes the bulk matter inside the cluster,
whereas the second describes the 
unbound neutron fluid in the
remaining fraction $1-u$ of the cell. 
The terms $\epsilon_e$, $\epsilon_{surf}$, and $\epsilon_C$ are
the energy density contributions of the electrons, surface,
and Coulomb interaction, respectively.

For spherical clusters with uniform charge densities, the
leading finite-size contributions are
\begin{eqnarray}
\epsilon_{\rm surf}
=\frac{3u\sigma_{\rm eff}}{R_N},
\nonumber \\
\epsilon_{\rm C}
=\frac{4\pi}{5}e^2(x_i n_i)^2R_N^2u
\left(1-\frac{3}{2}u^{1/3}+\frac{1}{2}u\right).
\end{eqnarray}
The effective surface tension $\sigma_{\rm eff}$ depends on
the properties of both the cluster and the surrounding neutron fluid. 
The Coulomb term includes the contribution of the
neutralizing electron background in the WS approximation.

At each $\bar n$, the cell energy is minimized under the
baryon-conservation and charge-neutrality constraints. The
pressure is obtained by the standard calculation:
\begin{equation}
P=\bar n^{\,2}\frac{d}{d\bar n}
\left(\frac{\epsilon_{\rm cell}}{\bar n}\right).
\end{equation}
The appearance of a nonzero $n_f$ marks neutron drip.

\subsection{BBP+Smooth Crossover}

In recent years, some studies have pointed out that there are problems with the direct connection proposed by Gledenning. Ref.~\cite{Fortin2016} made a comparison between the BPS+BBP crust EOS and a unified EOS, revealing differences in the radius of a canonical neutron star ranging from 60 to 150 m. For a stellar radius of 13 km, this corresponds to an uncertainty of approximately 1\%, which can be considered a small issue. It is also worth pointing out that unified EOSs predict the crust-core phase transition at
higher densities, around $0.08$ fm$^{-3}$. A more serious issue is related to the dimensionless tidal deformability, $\Lambda$. It was pointed out in Ref.~\cite{Pang2026PRD} that a direct connection can lead to an erroneous estimate of $\Lambda$, with deviations reaching up to 165.

Therefore, a second approach is to use the same BBP EOS up to 0.01 fm$^{-3}$ and then connect it to the core EOS through a smooth crossover, as suggested in Ref.~\cite{Pang2026PRD}:

\begin{equation}
\begin{aligned}
\epsilon(P) ={}&
\frac{1}{2}
\left[
1+\tanh\left(\frac{P-P_C}{\Delta P}\right)
\right]\epsilon_{\rm core}(P)
\\
&+
\frac{1}{2}
\left[
1-\tanh\left(\frac{P-P_C}{\Delta P}\right)
\right]\epsilon_{\rm crust}(P).
\end{aligned}
\label{eq:crossover}
\end{equation}

Using the crust and core EOSs to provide the initial conditions.  To ensure thermodynamic consistency, the baryon number density $n(P)$ is obtained by integrating the following differential equation~\cite{Haensel2004}:

\begin{equation}
\frac{dn}{dP}
=
\frac{n(P)}{\epsilon(P)+P}
\frac{d\epsilon(P)}{dP},
\label{eq:thermodynamic_consistency}
\end{equation}

The interpolation procedure is applied iteratively until the step in the density becomes lower than $2\times10^{-3}$ fm$^{-3}$ at both ends of the matching region. Of course, there is an ambiguity about the value of $\epsilon_{\rm core}$. Ref.~\cite{Pang2026PRD} studied the smooth crossover assuming $P_{cc}$ = 0.25 and  0.30 MeV/fm$^{3}$ without found any significant difference. On the other hand, Ref.~\cite{Xu2009PRC} suggests that the crust--core phase transition happens in the region  0.040 fm$^{-3}< n <$ 0.065 fm$^{-3}$ with  0.01 MeV/fm$^{3}< P_{cc} <$ 0.26 MeV/fm$^{-3}$, and, as pointed out earlier, Ref.~\cite{Fortin2016} points to a crust--core phase transition around 0.08 fm$^{3}$.

Therefore, to avoid large deviations in these predictions,
we determine $\epsilon_{\rm core}$ in such a way that the first point in the core EOS  simultaneously satisfies $n>0.06$ fm$^{-3}$ and $P_{cc}>0.25$ MeV/fm$^{3}$.

\begin{table*}[ht!]
\centering
\caption{Crust-core transition properties of the CLDM crust for the four
symmetry-energy slopes $L$ studied, computed with the CUTER V2 algorithm.
$B/A=-16.2$, $n_{\rm 0}=0.156$, $K_{\rm 0}=256$, 
$Q_{\rm sat}=-353$, $Z_{\rm sat}=2835$, $S_{\rm 0}=31.7$ MeV).}
\label{tab:crust_core_transition}
\begin{tabular}{ccccccccccc}
\hline
$L$ (MeV) & $K_{\rm sym}$ (MeV) & $Q_{\rm sym}$ (MeV) &
$n_{\rm drip}$ (fm$^{-3}$) &
$n_{\rm cc}$ (fm$^{-3}$) &
$P_{\rm cc}$ (MeV\,fm$^{-3}$) &
$Y_{p(CUTER)}$ & $Y_{p(QHD)}$ & $\Delta Y_p$\\
\hline
44 & $-108$ & 1384 & $2.49 \times10^{-4}$ & $~8.75\times10^{-2}$ & 0.46 & $3.88\times10^{-2}$ & $3.80\times10^{-2}$ &  $~+0.8\times10^{-3}$\\
60 & $-129$  & 777  &$2.59\times10^{-4}$ & $~7.97\times10^{-2}$ &  0.53 & $2.81\times10^{-2}$ & $2.71\times10^{-2}$ & $+~1.0\times10^{-3}$ \\
76 & $-84$   &140  & $2.86\times10^{-4}$ & $~6.57\times10^{-2}$ & 0.33 &$1.80\times10^{-2}$ & $1.74\times10^{-2}$ & $+~0.6\times10^{-3}$  \\
92 & $+27$   &80   & $3.42\times10^{-4}$ & $~5.76\times10^{-2}$ &  0.18 &$1.29\times10^{-2}$ & $1.18\times10^{-2}$ & $+~1.1\times10^{-3}$ \\
\hline
\end{tabular}
\end{table*}

\subsection{Unified EOS}

An ideal description of the crust is provided by a unified crust--core EOS, in which the same underlying nuclear matter properties are consistently employed throughout the star. In our approach, homogeneous core matter was described within QHD, while clustered matter in the crust is treated within the CLDM. The consistency between these two descriptions is ensured by using the same nuclear parameters, which determine the bulk properties of nuclear clusters and the surrounding neutron fluid in the inner crust. Therefore, these parameters are not independently adjusted in the crust model, but are inherited from the QHD parametrization.

For any nuclear matter, the energy per baryon of the homogeneous phase can be
separated into an isoscalar contribution, corresponding to
symmetric matter ($Z=N$), and an isovector contribution
associated with the neutron--proton asymmetry ($Z\neq N$). Within the
parabolic approximation, and using our notation, we
write~\cite{Margueron2018a,Pieka2009PRC}:

\begin{equation}
\epsilon
=
\epsilon_0(n)
+\alpha^2 S(n) +\mathcal{O}(\alpha^4), \quad \mbox{with}
\label{eq:energy_per_baryon}
\end{equation}

\begin{equation}
\begin{aligned}
\epsilon_0(n) ={}&
B/A
+\frac{1}{2}K_0 x^2
+\frac{1}{3!}Q_{\rm sat}x^3\\
&+\frac{1}{4!}Z_{\rm sat}x^4+\cdots,
\end{aligned}
\label{eq:isoscalar}
\end{equation}

\begin{equation}
\begin{aligned}
S(n) ={}&
S_{\rm 0}
+Lx
+\frac{1}{2}K_{\rm sym}x^2\\
&+\frac{1}{3!}Q_{\rm sym}x^3
+\cdots, \quad \mbox{where}
\end{aligned}
\label{eq:isovector}
\end{equation}

\begin{equation}
x=\frac{n-n_{\rm 0}}{3n_0},
\qquad
\alpha=\frac{n_n-n_p}{n_n+n_p}.
\end{equation}

The isoscalar coefficients are evaluated at the saturation
density $n_0$:
\begin{equation}
\begin{aligned}
K_0 &= 9n_0^2
\left.\frac{d^2\epsilon_0}{dn^2}\right|_{n_0},\\
Q_{\rm sat} &= 27n_0^3
\left.\frac{d^3\epsilon_0}{dn^3}\right|_{n_0},\\
Z_{\rm sat} &= 81n_0^4
\left.\frac{d^4\epsilon_0}{dn^4}\right|_{n_0}.
\end{aligned}
\end{equation}

Likewise, the symmetry-energy coefficients are
\begin{equation}
\begin{aligned}
S_0 &=
\left.\frac{1}{2}
\frac{\partial^2\epsilon(n,\alpha)}
     {\partial\alpha^2}\right|_{n=n_0,\,\alpha=0},\\
L &= 3n_0
\left.\frac{dS}{dn}\right|_{n_0},\\
K_{\rm sym} &= 9n_0^2
\left.\frac{d^2S}{dn^2}\right|_{n_0},\\
Q_{\rm sym} &= 27n_0^3
\left.\frac{d^3S}{dn^3}\right|_{n_0}.
\end{aligned}
\end{equation}

The unified crust EOS employed in this work is obtained with the
publicly available \textsc{CUTER} v2 tool~\cite{Davis2024,Davis2025}.
As input, we provide the EOS of homogeneous matter in
$\beta$ equilibrium calculated with QHD, together with its
nuclear matter parameters at saturation, presented in Tab~\ref{TL1} and Tab.~\ref{TL2}.~\textsc{CUTER} v2 uses these
inputs to reconstruct the nuclear matter functional at
sub-saturation densities and calculate the crust EOS within
an improved version of the CLDM, once in addition to the terms related to finite size, such as the Coulomb and surface terms, it also takes into account curvature contributions
at leading order. Furthermore, it uses the updated nuclear chart from the AME2020 mass table~\cite{Wang2021} to adjust the CLDM parameters, rather than the now outdated version from the early 1970s used in the BBP EOS. Additional details can be seen in Refs.\cite{Davis2024,Davis2025,Pieka2009PRC,Margueron2018a}.

The original characterization of the QHD model was limited to
the lower-order properties of nuclear matter at saturation.
Consequently, the higher-order coefficients $Q_{\rm sat}$,
$Z_{\rm sat}$, $K_{\rm sym}$, and $Q_{\rm sym}$ were not
previously reported. Here, we calculate them directly from
the QHD energy per baryon using the derivatives defined above.
In particular, the resulting isoscalar and isovector coefficients are used
in the reconstruction of the crust with \textsc{CUTER} v2.
Together with the QHD EOS in $\beta$ equilibrium, this more
complete characterization of nuclear matter allows for a consistent description of NS crust and core.

The parameters $Q_{\rm sat}$ and $Z_{\rm sat}$ are related solely
to the isoscalar part, and therefore have the same value for all 
$L$ analyzed. We obtain $Q_{\rm sat} = $ -353 MeV and $Z_{\rm sat} =$ +2835 MeV.
On the other hand, the isovector parameters depend on the $\rho$ field and therefore vary with the slope.  We present their values in Tab.~\ref{tab:crust_core_transition}.

The \textsc{CUTER} v2 EOS begins at 10$^{-4}$ fm$^{-3}$ for all values of $L$ (for lower densities the BPS EOS is used).
We also define the crust--core phase transition criterion as the density at which the homogeneous phase becomes energetically favorable over the clusterized one. Explicitly:

\begin{equation}
  \bigg (\frac{E}{A}  \bigg )_{\mbox{hom}} <  \bigg ( \frac{E}{A}  \bigg )_{\mbox{crust}}
\end{equation}

%%%%%%%%%%%%%%%%%%%%%%%%%%%%%%%%%%%%%%%%%%%%%%%%

\section{Numerical Results I: Microphysics} \label{sec:micro}

\textsc{CUTER} v2 allows us to study properties coming directly from the NS crust. We can begin analyzing how the slope affects neutron drip. We found that the neutron drip point grows monotonically with the slope, varying from $2.49 \times10^{-4}$ fm$^{-3}$ for $L = 44$ MeV, up to $3.42 \times10^{-4}$ fm$^{-3}$ for $L = 92$ MeV. 
For comparison, the BBP EOS predicts $n_{\rm drip}=2.59\times10^{-4}$ fm$^{-3}$~\cite{BBP}, exactly the same value predicted for $L=60$ MeV. An even more important physical quantity also shows a monotonic behavior with the slope: the crust--core density transition, $n_{cc}$.  Here, however, the effect is reversed. The $n_{cc}$ decreases with the slope. For $L = 44$ MeV we have $n_{cc}=8.75\times10^{-2}$ fm$^{-3}$, while for $L = 92$ MeV, $n_{cc}=5.76\times10^{-2}$ fm$^{-3}$. The pressure at the crust--core transition, $P_{cc}$, shows non-monotonic behavior, with a maximum at $L = 60$ MeV. We cannot compare our results of the crust--core transition with the BBP EOS because the authors use a different criterion. Nevertheless, for most values of $L$, we find $P_{cc}$ higher than those discussed in Refs.~\cite{Pang2026PRD,Xu2009PRC}, although the literature reports values as high as 0.76 MeV/fm$^3$~\cite{Moustakidis2010}. The complete results are displayed in Tab.~\ref{tab:crust_core_transition}.

%%%%%%%%%%%%%%%%%
\begin{figure*}[ht]
\begin{tabular}{ccc}
\centering % \begin{center}/\end{center} takes some additional vertical space
\includegraphics[scale=.58, angle=270]{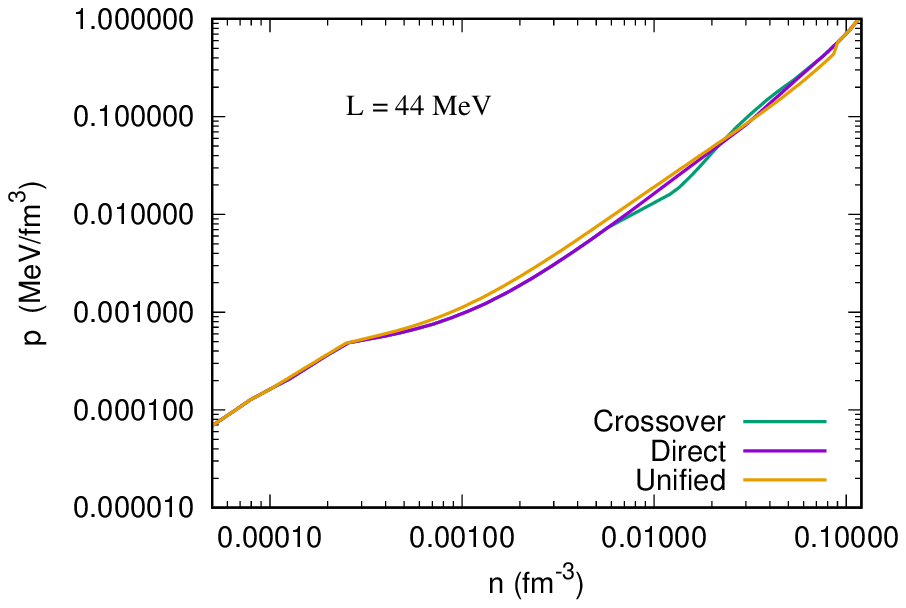} &
\includegraphics[scale=.58, angle=270]{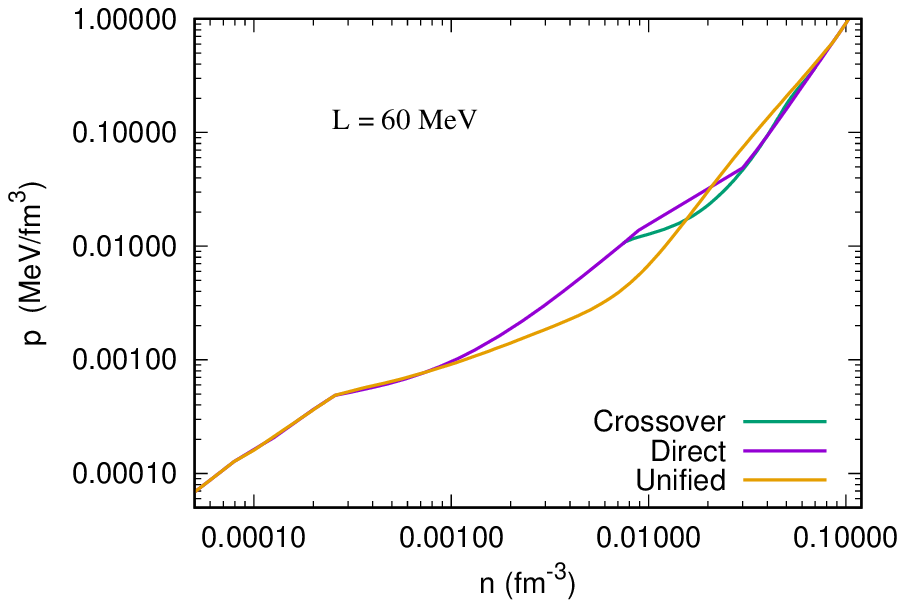} \\
\includegraphics[scale=.58, angle=270]{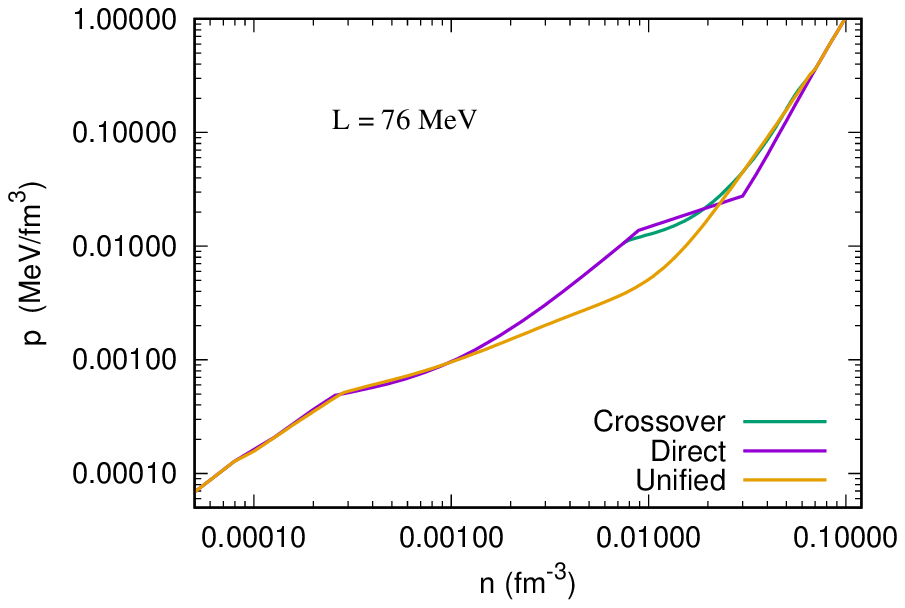} &
\includegraphics[scale=.58, angle=270]{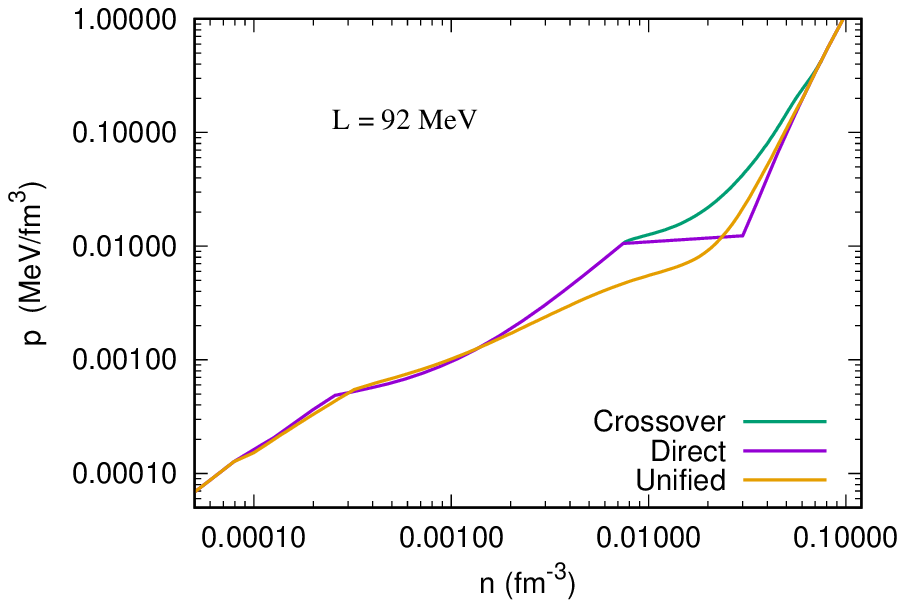}\\
\end{tabular}
\caption{The crust EOSs for different values of the slope $(L)$ within the three disucssed approach} \label{CrustEOS}
\end{figure*}

\begin{figure}[ht]
  \begin{centering}
\begin{tabular}{c}
\includegraphics[width=0.334\textwidth,angle=270]{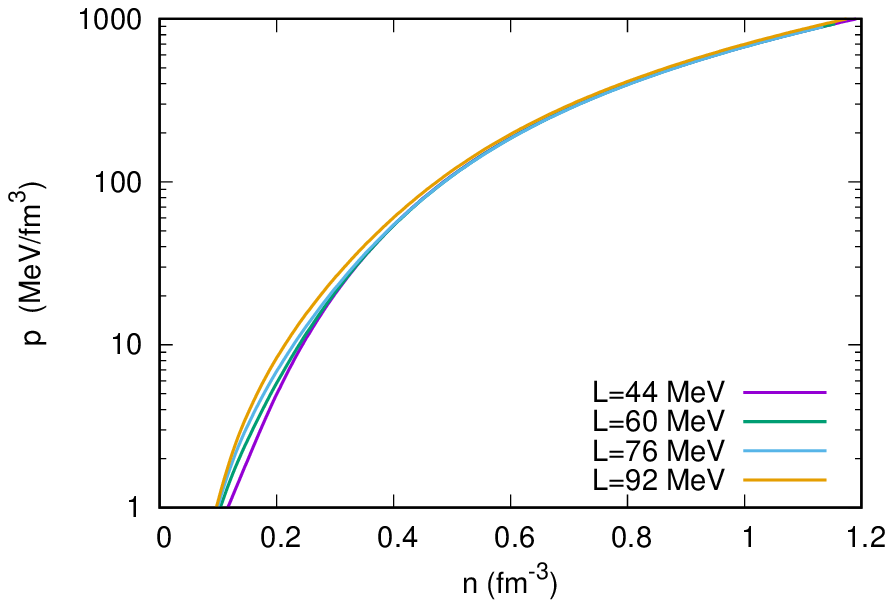} \\
\end{tabular}
\caption{The core EOS  different values of $L$.} \label{coreEOS}
\end{centering}
\end{figure}

We now turn our attention to the equations of state. In Fig.~\ref{CrustEOS} we display the EOSs for the crust within the three approaches and for the four values of $L$.
Beginning with $L = 44$ MeV, we see that for this particular value of $L$ there are almost no differences between the EOSs. The unified EOS predicts a slightly stiffer EOS compared with the direct connection, while the smooth crossover causes a small concavity, softening the EOS. For this value of $L$, there is no visible discontinuity in the direct connection as well.

For $L = 60$ MeV, there is already a displacement of the unified EOS when compared with the direct connection. The smooth crossover also produces a more pronounced concavity. The discontinuity in the direct connection begins to be visible. Qualitatively, the results are very similar for $L = 76$ MeV.

$L = 92$ MeV presents the situation where the three approaches present the largest differences among them. The direct connection presents a clear discontinuity, represented as an almost horizontal step. The smooth crossover links the crust to the core, producing a clearly stiffer EOS, while the unified EOS is essentially softer than the direct connection before being connected to the core.

We can also investigate the reverse situation. For a fixed approach, what is the effect of the slope in the crust EOS? We found that for all three approaches, larger values of $L$ produce a softer EOS, until around 0.08 fm$^{-3}$, then the situation is reversed. 
This behaviour has already been noted in Ref.~\cite{lopescesar} for the direct connection. 

In Fig.~\ref{coreEOS}, we display the core EOSs for different values of $L$ for $P>1$ MeV/fm$^3$. In this region, the stiffening has already inverted, and large $L$ values produce stiffer EOSs. However, we can notice that for $n~\approx 0.4$ fm$^{-3}$, the results become essentially degenerate.

\begin{figure}[ht]
  \begin{centering}
\begin{tabular}{c}
\includegraphics[width=0.334\textwidth,angle=270]{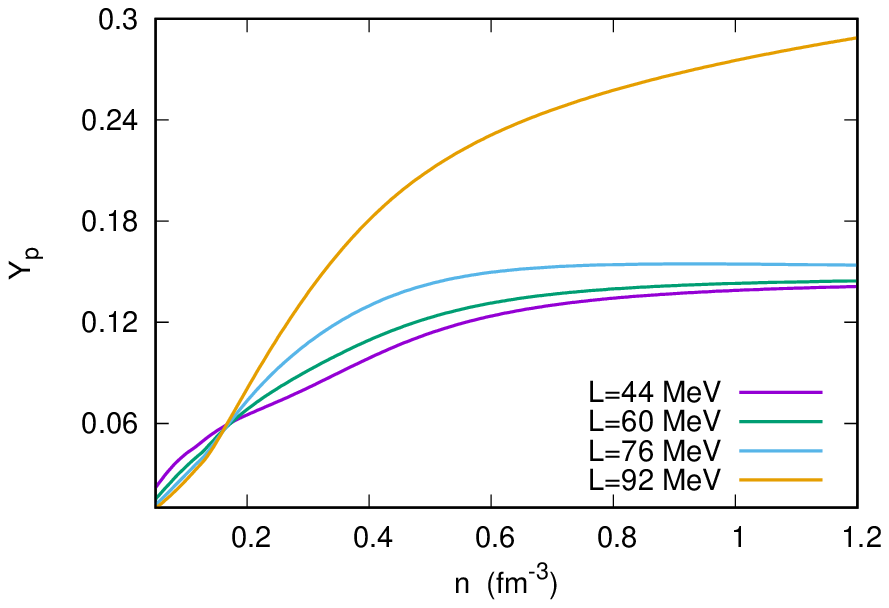} \\
\includegraphics[width=0.334\textwidth,angle=270]{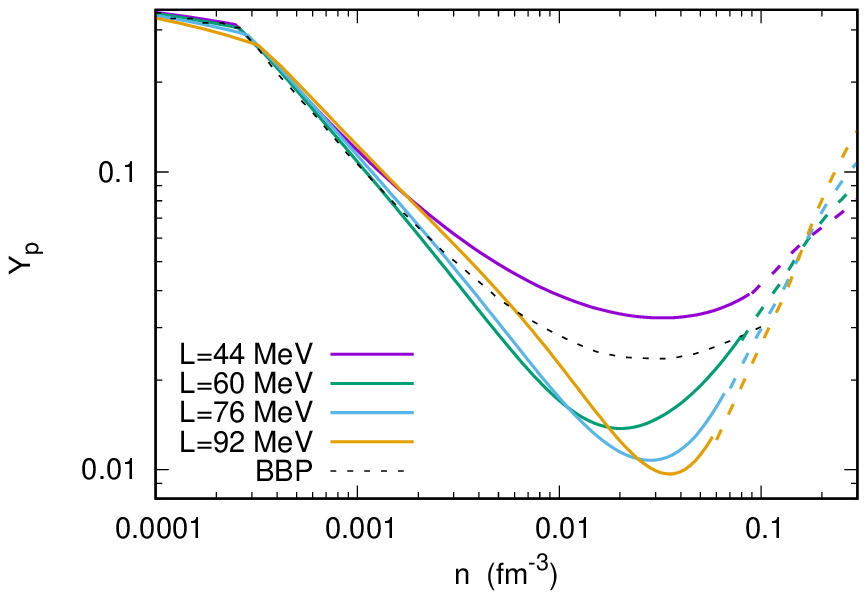}
\end{tabular}
\caption{The proton fraction for the core (above) and the crust (below) for different values of $L$. The BBP proton fraction is also displayed for the crust for comparison.} \label{protonfraction}
\end{centering}
\end{figure}

Another important feature is the proton fraction, $Y_p$, which is linked to the behaviour of the symmetry energy of nuclear matter, as well as to the Direct URCA process in the stars~\cite{Lattimer1991,lopes2024PRCb}. Fig.~\ref {protonfraction} presents the results for the core and the crust.

Beginning with the core, we see that, as happened in the analysis of the pressure, the effects of $L$ are opposite in the low- and high-density regimes. For densities up to 0.18 fm$^{-3}$, large values of $L$ produce lower values of the proton fraction. For densities above this point, the result is inverted. In the high-density limit, $Y_p$ reaches around to 0.27 for $L = 92$ MeV, but only 0.15 for $L = 44$ MeV. 
We observe more complex behavior in the crust because different physical effects, such as surface and Coulomb contributions, compete, whereas they are absent in homogeneous core matter. At $n=10^{-4}$ fm$^{-3}$, the proton fraction is around 0.35 and decreases smoothly up to the neutron drip. After the neutron drip, we observe a sharp decrease, reaching values below 0.04. After $Y_p$ reaches a minimum close to $n=0.03$ fm$^{-3}$, it gradually increases until the crust--core transition density, $n_{cc}$, is reached. 

We can also investigate how robust \textsc{CUTER} v2 is by comparing $Y_p$ at $n_{cc}$ predicted by the CLDM and the QHD. The results coming from the CLDM are displayed at the bottom of Fig.~\ref{protonfraction} as solid lines up to $n_{cc}$, while those coming from the QHD are displayed as dashed lines beginning at $n_{cc}$. Qualitatively, the results are very well behaved. At $n_{cc}$, the lower the value of $L$, the larger $Y_p$ is, fully consistent with the core behaviour. The quantitative results are presented in Tab.~\ref{tab:crust_core_transition}. It can be seen that the differences barely reach $10^{-3}$. The largest difference is found for  $L = 92$ MeV, yet it is still only $1.1\times 10^{-3}$. Therefore, for all values of $L$, the differences between the CLDM and the QHD are negligible at $n_{cc}$.

For matter of completeness, we also present the proton fraction for the BBP EOS~\cite{BBP}. Unsurprisingly, the qualitative behaviour is the same as that obtained by the \textsc{CUTER} v2, once both models are derived from the CLDM.

\section{Numerical Results II: Macrophysics} \label{sec:macro}

We now turn our attention to the macroscopic properties of NS within the three discussed approaches to the crust and the four values of the slope. We divide this section into two parts: one related to the global properties of NS, i.e., the NS as a whole, and one related exclusively to the crustal properties, considering only the values coming from the crust. 

\subsection{Global properties}

%%%%%%%%%%%%%%%%%
\begin{figure*}[ht]
\begin{tabular}{ccc}
\centering % \begin{center}/\end{center} takes some additional vertical space
\includegraphics[scale=.54, angle=270]{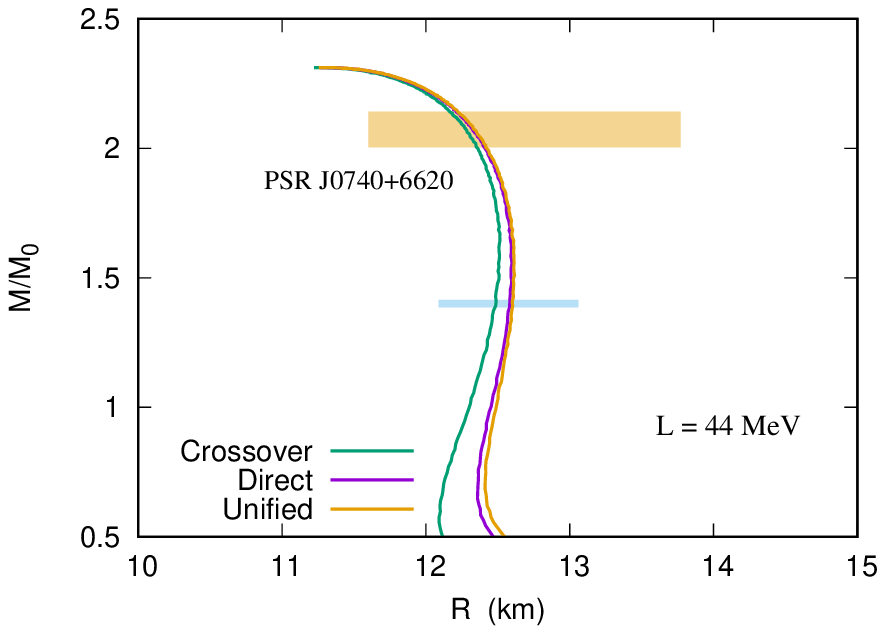} &
\includegraphics[scale=.54, angle=270]{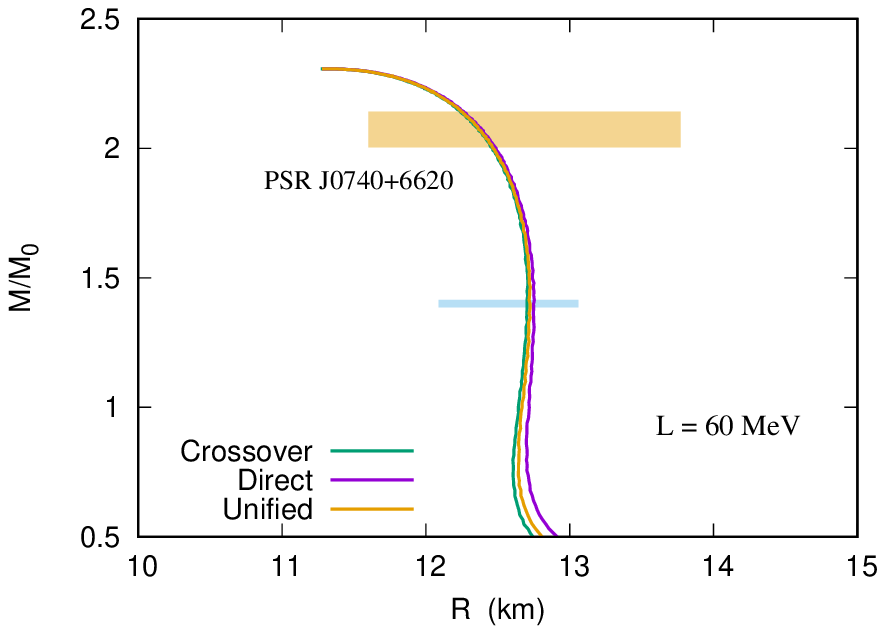} \\
\includegraphics[scale=.54, angle=270]{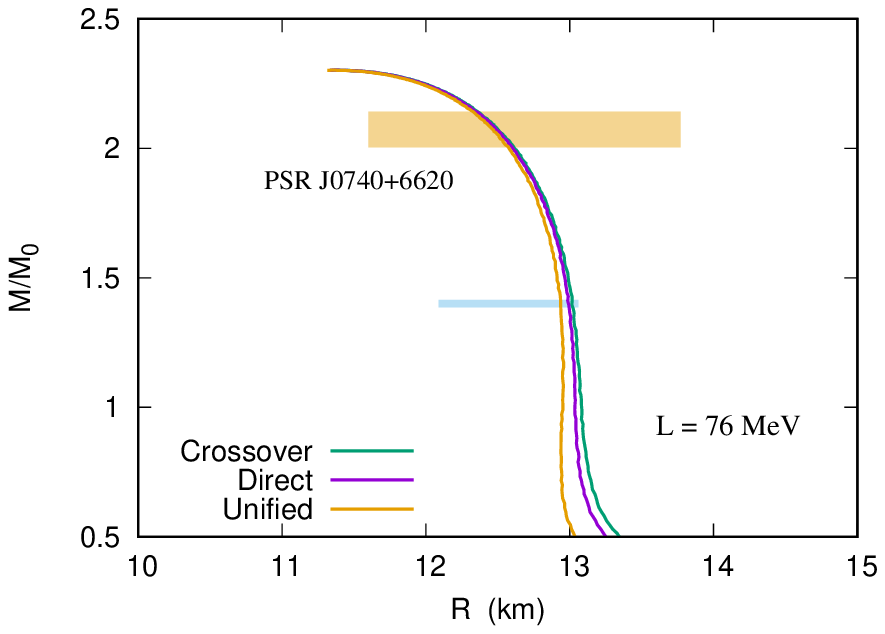} &
\includegraphics[scale=.54, angle=270]{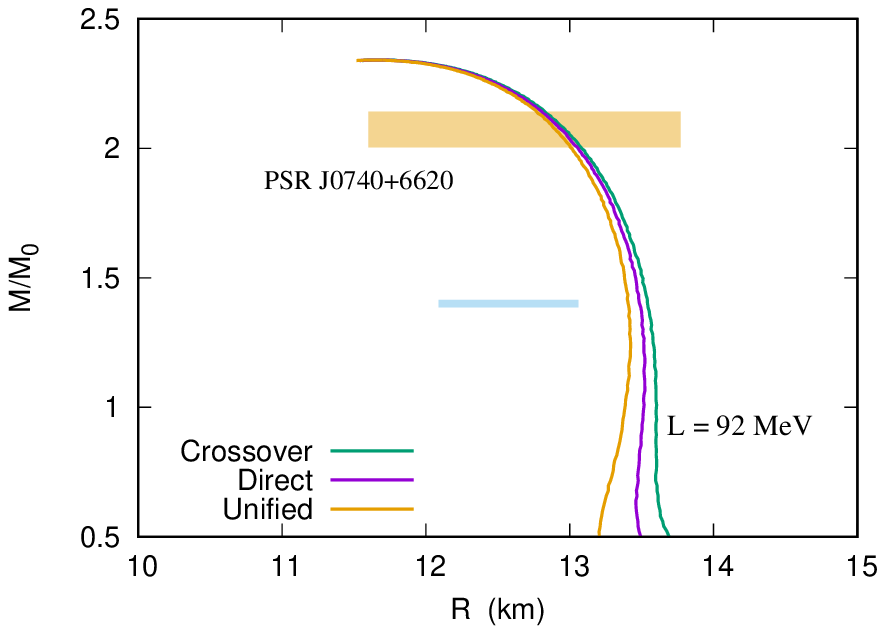}\\
\end{tabular}
\caption{Mass-radius relation for different values of the slope $(L)$ within the three discussed approaches. The hatched areas are the constraints discussed in the text related to the PSR J0740+6620~\cite{Salmi_2024}, and the canonical 1.4 $M_\odot$ pulsar~\cite{Miller_2026}.} \label{tovs}
\end{figure*}

We begin by analysing the mass-radius relation from the solution of the Tolman-Oppenheimer-Volkoff (TOV) equations~\cite{TOV}:

\begin{eqnarray}
 \frac{dp}{dr}
 =
 -\frac{GM(r)\epsilon(r)}{r^2}
 \bigg [\left(
 1+\frac{p}{\epsilon}
 \right)
 \left(
 1+\frac{4\pi p r^3}{M(r)}
 \right) \nonumber \\
 \times \left(
 1-\frac{2GM(r)}{r}
 \right)^{-1}\bigg ]. \nonumber
 \end{eqnarray}

\begin{equation}
M(r)
 =
\int_0^r 4\pi r^2\epsilon(r') \, d^3r'
 \label{OV}
\end{equation}
with the initial conditions $p(0) = p_0$ and $M(0) = 0$. The boundary condition $p(R) = 0$ defines the surface of the star. Therefore, for a given EOS, each value of $p_0$ represents a different star. The TOV solution for each value of $L$ and crust approach is presented in Fig.~\ref{tovs}.

Altogether, we also present two modern observational constraints coming from the NICER X-ray telescope~\cite{NICER2012}. In Ref.~\cite{Salmi_2024}, the authors inferred the mass and radius of the PSR J0740+6620 with 3.6 years of NICER Data. They report $M = 2.073 \pm 0.069~M_\odot$, and $R = 12.49^{+1.28}_{-0.88}$ km. In the same sense, Ref.~\cite{Miller_2026} updated the constraint on the radius of the canonical 1.4 $M_\odot$ NS to lie in the range 12.56 $\pm$ 0.47 km, implying an uncertainty of only 4$\%$; both constraints are at 68\% credible interval. We can see in Fig.~\ref{tovs} that the well-known effect of the increase of the radii as we increase the slope~\cite{Rafa2011,lopescesar} is also present here. Furthermore, it is clear that $L = 92$ MeV produces radii too large, in disagreement with the constraint related to the canonical 1.4 $M_\odot$.

Now, considering the unified EOS as the best description for the crust, we can ask how the different approaches compare with it. For $L = 44$ MeV, both predict lower values for the radii close to the canonical star, while for $L = 76$ and 92 MeV, both predict larger values when compared with the unified EOS. $L = 60$ MeV has a more curious behaviour, with the direct connection slightly larger and the smooth crossover slightly lower than the unified scheme. For all values of $L$, the differences between the direct connection and the unified EOS are around $1\%$ for the canonical star, in agreement with what was discussed in Ref.~\cite{Fortin2016}. Moreover, smooth crossover, in general, produces larger variations when compared with the direct connection. Therefore, for the mass-radius relation, the direct connection is preferable to the smooth crossover. 
The quantitative results for stars from 1.2$M_\odot$ up to 2.0$M_\odot$ for all values of $L$ are presented in Tab.~\ref{tab:crust_direct} for the direct connection, Tab~\ref{tab:cross} for the smooth crossover, and Tab.~\ref{tab:crust_properties_unified} for the Unified EOS.

\begin{figure}[ht]
  \begin{centering}
\begin{tabular}{c}
\includegraphics[width=0.334\textwidth,angle=270]{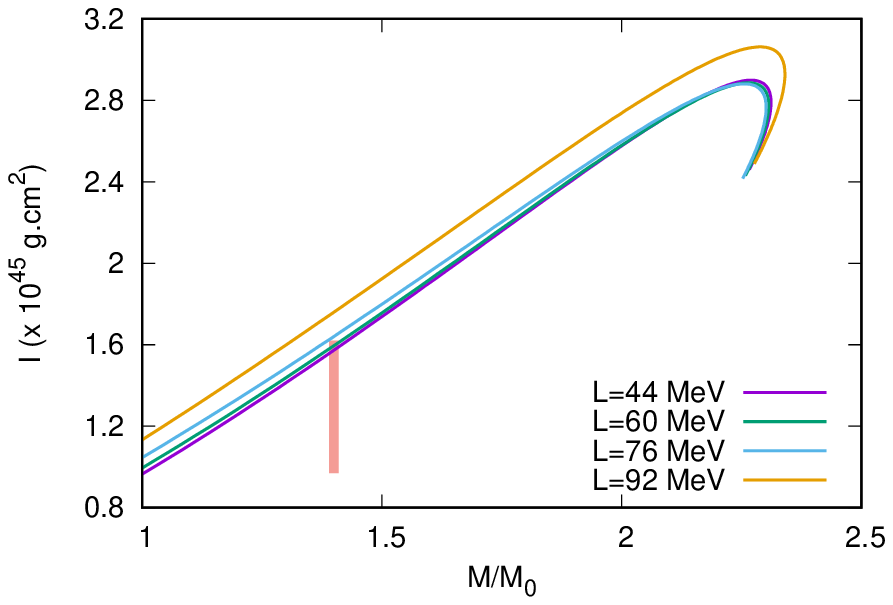} \\
\end{tabular}
\caption{MOI for different values of $L$ in the unified scheme and the inferred constraint $I_{1.4} = 1.22^{+0.40}_{-0.25}$ g.cm$^2$ presented in Ref.~\cite{Wang2022PRC}. There are no significant differences between different crust approaches.} \label{MOI}
\end{centering}
\end{figure}

We now analyze the moment of inertia (MOI) for four values of $L$. As can be seen in Tab.~\ref{tab:crust_direct}, Tab.~\ref {tab:cross}, and Tab.~\ref{tab:crust_properties_unified}, the effect of the crust is negligible. Therefore, we only discuss the effect of the slope assuming a unified EOS.
The MOI can be calculated from~\cite{Glenbook}:

\begin{equation}
I =
\frac{8\pi}{3}
\int_0^R
\frac{
r^4\left(\epsilon+P\right)
}{
\sqrt{1-\dfrac{2GM(r)}{r}}
}
e^{-\nu(r)/2}
\frac{\bar{\omega}(r)}{\Omega}
\,dr , \label{moi}
\end{equation}
with:

\begin{equation}
\frac{d\nu}{dr}
=
\frac{2G}{r^2}\,
\frac{M(r)+4\pi r^3P(r)}
{\left[1-\dfrac{2GM(r)}{r}\right]}.
\end{equation}

The results are presented in Fig.~\ref{MOI} altogether with the constraint $I_{1.4} = 1.22^{+0.40}_{-0.25}$ g.cm$^2$ presented in Ref.~\cite{Wang2022PRC}. Now, this is not a direct observational constraint, but an inferred constraint obtained by combining the tidal deformability from GW170817 and the $I-Love-Q$ universal relations from relativistic $ab~initio$ methods. As can be seen, MOI increases with the slope, and  $L = 76$ MeV already produces values too large for $I_{1.4}$. $L = 92$ MeV produces a particullary large value of $I_{1.4}$,due to its large radius, $R_{1.4} = 13.41$ km.

Indeed, the large increase in the radius for $L = 92$ MeV cannot be justified only by the value of the slope, since $\Delta L$ = 16 MeV in all cases. But also for the value of $K_{sym}$. $L = 92$ MeV is the only parametrization that produces a positive value of $K_{sym} = +27$ MeV. Furthermore, it also has a large increase when compared with $L = 76$ MeV, which has $K_{sym} = -87$ MeV. This implies that large orders in the symmetry energy expansion can also be important. 
For even more massive stars, $K_{sym}$ becomes even more important. For a $2.0~M_\odot$, we see that $I_{2.0} = 2.59 \pm 0.01$ g.cm$^2$ for $L = 44,~60$ and 76 MeV, but increases to 2.75 g.cm$^2$ for $L = 92$ MeV.

\begin{figure}[ht]
  \begin{centering}
\begin{tabular}{c}
\includegraphics[width=0.311\textwidth,angle=270]{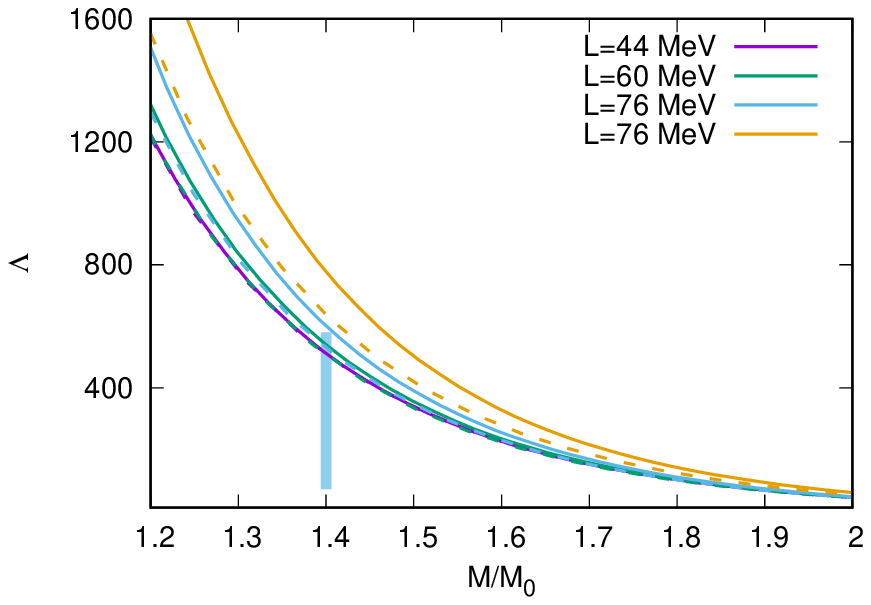} \\
\includegraphics[width=0.311\textwidth,angle=270]{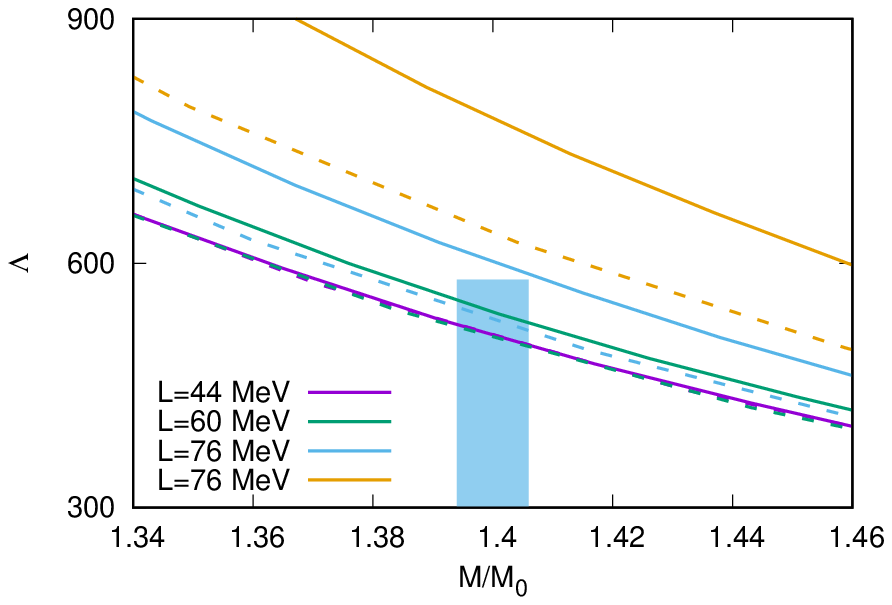}
\end{tabular}
\caption{(Above) dimensionless tidal parameter $\Lambda$ for different values of $L$ in the unified (solid lines) and the direct connection (dashed lines). (Below) Zoom in around 1.4$M_\odot$. The hatched area is the constraint coming from the GW170817 event, $70<\Lambda_{1.4}<580$ as discussed in Ref.~\cite{AbbottPRL}. Differences between the unified and the crossover schemes are negligible.} \label{tidal}
\end{centering}
\end{figure}

The last but not least global property analyzed here is the dimensionless tidal parameter $\Lambda$.  If we put an extended body in an inhomogeneous external field, it will experience different forces throughout its surface. The result is a tidal interaction. The tidal deformability of a compact object is a single parameter $\lambda$ that quantifies how easily the object is deformed when subjected to an external tidal field. Larger tidal deformability indicates that the object is easily deformed. Conversely, a compact object with a small tidal deformability parameter is more compact and more difficult to deform. The tidal deformability is defined as:
%%%%%%%%%%%%%%%%
\begin{equation}
 \Lambda~\equiv~\frac{\lambda}{M^5}~\equiv~\frac{2k_2}{3C^5} , \label{tidal}
\end{equation}
%%%%%%%%%%%%%%%
where $M$ is the mass of the compact object and $C = GM/R$ is its compactness. The parameter $k_2$ is called the second (order) Love number. Additional discussion about the theory of tidal deformability and the tidal Love numbers is beyond this work's scope and can be found in Refs.\cite{Lopes2022ApJ,Hinderer_2008,tidal2010,AbbottPRL,Chat2020,Flores2020}.
 
The study of the dimensionless tidal parameter is particularly relevant here since Ref.~\cite{Pang2026PRD} pointed out that a direct connection can lead to an incorrect estimate of it, producing deviations as high as $20\%$. The numerical results are displayed in Fig.~\ref{tidal} together with the constraint related to the canonical star, $70<\Lambda_{1.4}<580$, obtained from the LIGO/VIRGO gravitational wave observatories in the GW170817 event~\cite{AbbottPRL}. 

First thing we must point out is that the crossover approach produces negligible differences when compared with the unified EOS, with deviations around or below $1\%$, in agreement with the discussion presented in Ref.~\cite{Pang2026PRD}. Therefore, results for smooth crossover are not shown in order to not saturate the figure. Another feature that also agrees with Ref.~\cite{Pang2026PRD} is the deviation of $\Lambda_{1.4}$ produced by the direct connection when compared with a unified EOS. Surprisingly, for $L = 44$ MeV, the results are very close and fall below 1$\%$, which can be associated with an absence of discontinuity in the EOS for this value of $L$, as shown in Fig.~\ref{CrustEOS}. However, as we increase the slope, we also increase the differences. It still reaches negligible values for $L = 60$ MeV, but already reaches  $12\%$ for $L = 76$ MeV. Finally, we obtain a $\Delta \Lambda_{1.4} = 21\%$ for $L = 92$ MeV. Furthermore, as in the case of the MOI, the constraint related to the GW170817 event cannot be satisfied for $L~\gtrsim 76$ MeV. This is a strong refinement of Ref.~\cite{lopes2024PRC}, which rules out $L~\gtrsim100$ MeV for the same parametrization.
We can also notice a huge jump in $\Lambda_{1.4}$ from $L = 76$ MeV to $L = 92$ MeV $(\Delta \Lambda_{1.4} = 176)$. This suggests a major role played by the $K_{\rm sym}$ in the dimensionless tidal parameter as well. The numerical values for some NS masses are presented in Tab.~\ref{tab:crust_direct}, Tab.~\ref{tab:cross}, and Tab.~\ref{tab:crust_properties_unified}.

%%%%%%%%%%%%%%%%%
\begin{figure*}[ht]
\begin{tabular}{ccc}
\centering % \begin{center}/\end{center} takes some additional vertical space
\includegraphics[scale=.54, angle=270]{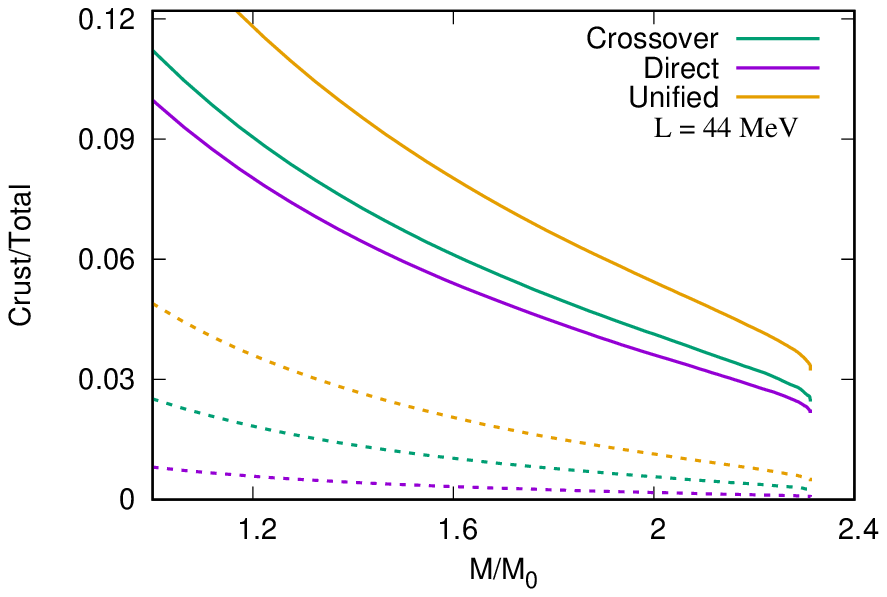} &
\includegraphics[scale=.54, angle=270]{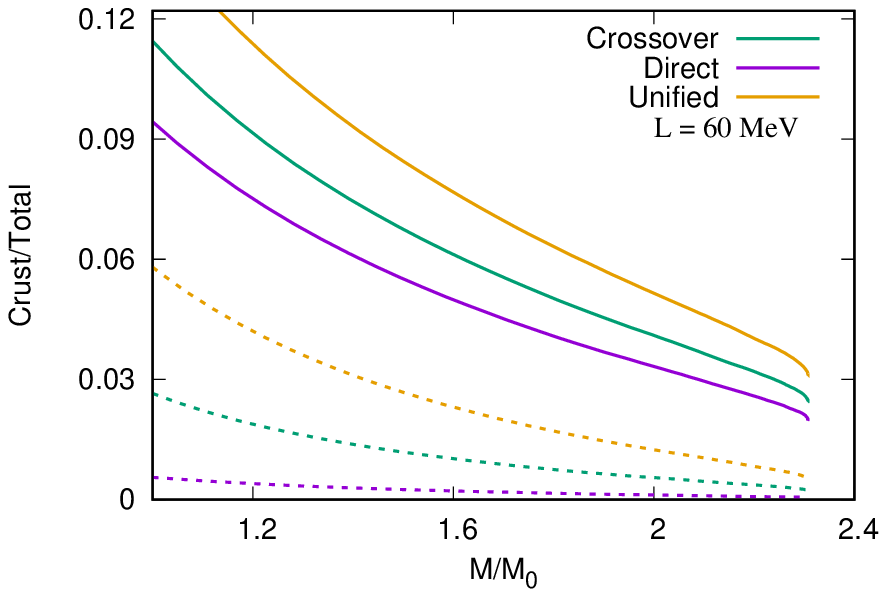} \\
\includegraphics[scale=.54, angle=270]{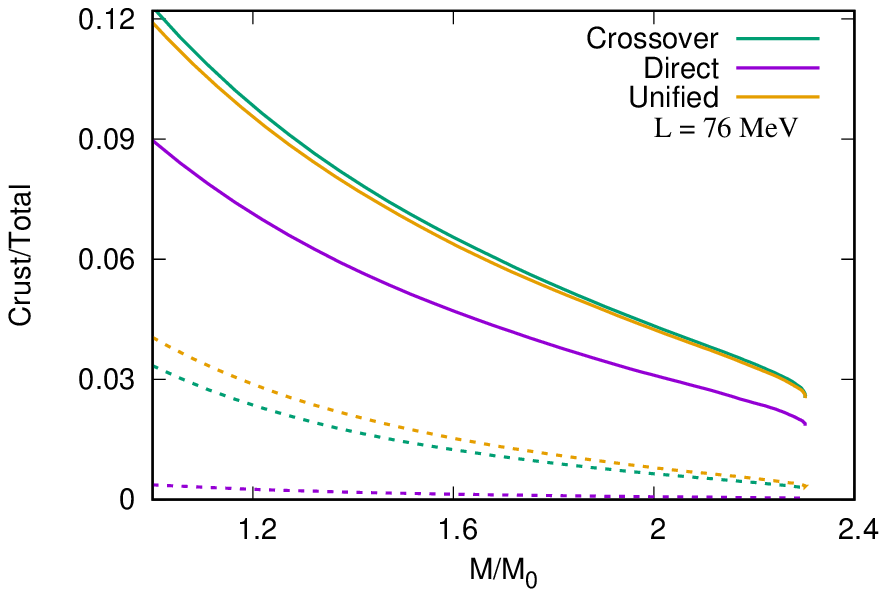} &
\includegraphics[scale=.54, angle=270]{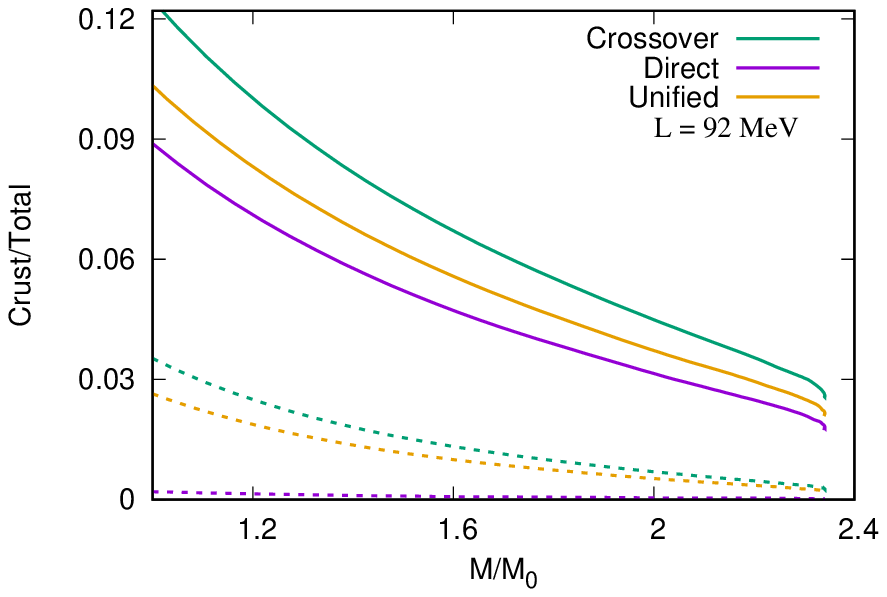}\\
\end{tabular}
\caption{Relative contribution of the crust for the radius (solid lines) and the mass (dashed lines) of NS for different values of $L$  within the three discussed approaches. } \label{relative}
\end{figure*}

We can conclude this section about the global properties, pointing out that in the absence of a unified treatment for the crust,  the direct connection gives us a satisfactory description of the mass-radius relation, as well as the MOI of the stars. However, in the calculation of the dimensionless tidal parameter, the smooth crossover is mandatory. And, as it predicts larger variations compared with the direct connection for the radius of the stars, it should be used solely for the calculation of $\Lambda$.

\subsection{Crustal Properties}

We can now discuss the effects of the slope and the three crust approaches specifically in the contribution of the crust in the total NS. Specifically, we calculate for a given EOS, what is the total and relative: size, mass and MOI of the crust.

To accomplish this task, we follow the procedure done in Ref.~\cite{lopes2024PRC} that calculates the mass and size of the quark core in hybrid stars. As pointed out earlier, for a given EOS, each value of central density $\epsilon_0$ corresponds to a different star. From the TOV equations, to calculate the core radius, we stop the integration at $P_{cc}$, so we have $p(R_{\rm core}) = P_{cc}$, which gives us the size of the core. The mass and MOI of the core are therefore obtained by replacing $R$ with $R_{\rm core}$ in Eq.~\ref{OV} and Eq.~\ref{moi}. Now we have:

\begin{eqnarray}
  M_{\rm crust} = M - M_{\rm core}, \quad R_{\rm crust} = R - R_{\rm core}, \nonumber
  \\ I_{\rm crust} = I - I_{\rm core},
\end{eqnarray}
while the relative contributions of the crust in relation to the total star are $M_{\rm crust}/M$, $R_{\rm crust}/R$, and $I_{\rm crust}/I$. The results for the mass (dashed lines) and radius (solid lines) are presented altogether in Fig.~\ref{relative}.

%%%%%%%%%%%%%%%%%
\begin{figure*}[ht]
\begin{tabular}{ccc}
\centering % \begin{center}/\end{center} takes some additional vertical space
\includegraphics[scale=.54, angle=270]{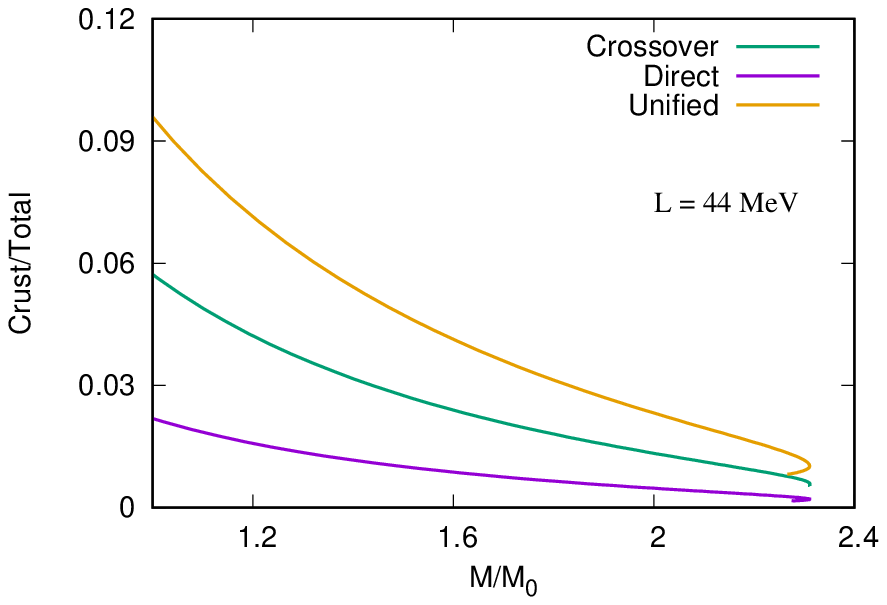} &
\includegraphics[scale=.54, angle=270]{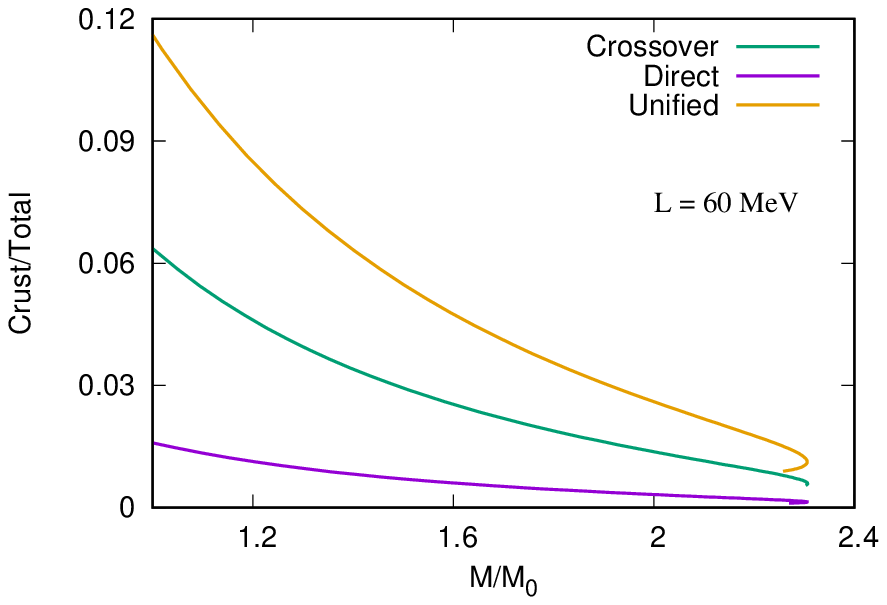} \\
\includegraphics[scale=.54, angle=270]{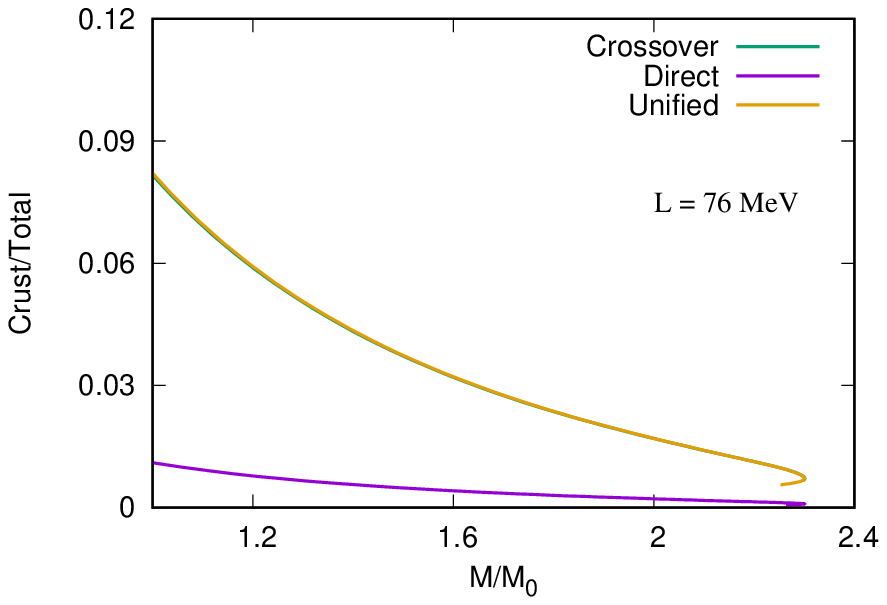} &
\includegraphics[scale=.54, angle=270]{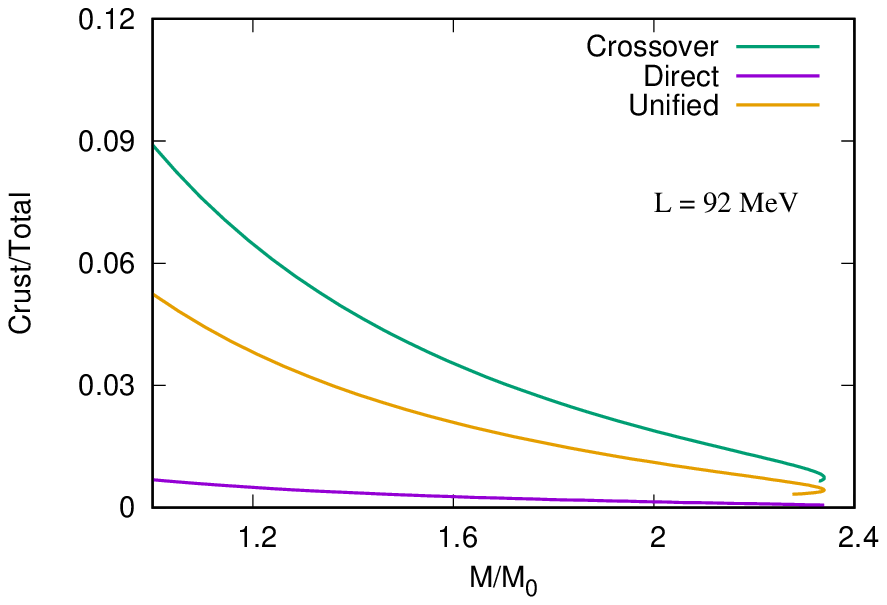}\\
\end{tabular}
\caption{Relative contribution of the crust for MOI of NS for different values of $L$  within the three discussed approaches.} \label{relativeMOI}
\end{figure*}

Let us begin with $M_{\rm crust}$, and specifically, in the direct connection scheme. This is, by far, the physical quantity that has the least influence on the star as a whole. As can be seen in Tab.~\ref{tab:crust_direct}, the crust mass in the direct connection barely reaches 0.5\%, and in some cases it drops below 0.1\%. In relation to the slope, we see that increasing $L$ reduces $M_{\rm crust}.$ In the same sense, the larger the star mass, the lower the crustal mass.

We now analyze the crust mass within the smooth crossover. The first feature we notice is a huge increase in $M_{\rm crust}$ for all stars analyzed; such an increase can surpass 200\%. In some cases, the crustal mass can reach values over 2\%. While the crustal mass still decreases with mass, the slope has the opposite effect. The larger the slope, the larger the crustal mass, as shown in Tab.~\ref{tab:cross}.

We now study the crustal mass within the unified scheme, which we believe best describes the crust. In most cases, the crustal mass is even larger than that predicted by the smooth crossover approach, and in some cases it can reach 4\% of the total mass. However, $L = 92$ MeV is an exception. This can be explained by analysing $n_{cc}$. $L = 92$ MeV is the only parametrization for which the unified EOS predicts a crust--core phase transition at a lower density than the smooth crossover. Another interesting feature is the dependence on the slope. In the unified approach, the crustal mass initially increases with $L$, reaching a maximum at $L = 60$ MeV, and then decreases monotonically. The numerical values are presented in Tab.~\ref{tab:crust_properties_unified}.

\begin{table*}[ht]
\centering
\caption{Global and crustal properties of neutron stars for different values 
of the symmetry energy slope ($L$). Direct connection.}
\label{tab:crust_direct}

\resizebox{\textwidth}{!}{%
\begin{tabular}{ccccccccccc}
\toprule
$L$ (MeV) &
$M/M_\odot$ &
$R$ (km) &
$I$ ($10^{45}\,\mathrm{g\,cm^2}$) &
$\Lambda$ &
$M_{\rm crust}$ ($M_\odot$) &
$M_{\rm crust}/M$ (\%) &
$R_{\rm crust}$ (km) &
$R_{\rm crust}/R$ (\%) &
$I_{\rm crust}$ ($10^{45}\,\mathrm{g\,cm^2}$) &
$I_{\rm crust}/I$ (\%) \\
\hline

44 & 1.2 & 12.53 & 1.26 & 1202 & 0.007 & 0.58 & 1.00 & 7.98 & 0.02 & 1.54 \\
   & 1.4 & 12.58 & 1.56 &  515 & 0.006 & 0.43 & 0.82 & 6.49 & 0.02 & 1.14 \\
   & 1.6 & 12.59 & 1.88 &  223 & 0.005 & 0.32 & 0.68 & 5.38 & 0.01 & 0.87 \\
   & 2.0 & 12.42 & 2.58 &   45 & 0.004 & 0.18 & 0.46 & 3.63 & 0.01 & 0.46 \\
\hline

60 & 1.2 & 12.74 & 1.27 & 1218 & 0.005 & 0.38 & 0.95 & 7.45 & 0.01 & 1.12 \\
   & 1.4 & 12.74 & 1.60 &  513 & 0.004 & 0.29 & 0.76 & 6.03 & 0.01 & 0.79 \\
   & 1.6 & 12.72 & 1.91 &  225 & 0.003 & 0.21 & 0.63 & 4.90 & 0.01 & 0.60 \\
   & 2.0 & 12.49 & 2.59 &   46 & 0.002 & 0.11 & 0.42 & 3.34 & 0.01 & 0.31 \\
\hline

76 & 1.2 & 13.02 & 1.33 & 1301 & 0.003 & 0.26 & 0.91 & 7.01 & 0.01 & 0.76 \\
   & 1.4 & 12.99 & 1.64 &  535 & 0.003 & 0.19 & 0.74 & 5.70 & 0.01 & 0.57 \\
   & 1.6 & 12.93 & 1.95 &  230 & 0.002 & 0.13 & 0.60 & 4.66 & 0.01 & 0.42 \\
   & 2.0 & 12.60 & 2.60 &   42 & 0.002 & 0.08 & 0.39 & 3.10 & 0.01 & 0.21 \\
\hline

92 & 1.2 & 13.52 & 1.44 & 1590 & 0.002 & 0.15 & 0.95 & 7.03 & 0.01 & 0.49 \\
   & 1.4 & 13.48 & 1.75 &  639 & 0.002 & 0.11 & 0.77 & 5.71 & $<0.01$ & 0.34 \\
   & 1.6 & 13.41 & 2.08 &  287 & 0.001 & 0.08 & 0.63 & 4.69 & $<0.01$& 0.25 \\
   & 2.0 & 13.06 & 2.75 &   58 & 0.001 & 0.05 & 0.41 & 3.17 & $<0.01$ & $0.14$ \\

\hline
\end{tabular}%
}
\end{table*}

\begin{table*}[ht]
\centering
\caption{Global and crustal properties of neutron stars for different values 
of the symmetry energy slope ($L$) within the smooth crossover region between the BBP and core equations of state.}
\label{tab:cross}

\resizebox{\textwidth}{!}{%
\begin{tabular}{ccccccccccc}
\hline
$L$ (MeV) &
$M/M_\odot$ &
$R$ (km) &
$I$ ($10^{45}\,\mathrm{g\,cm^2}$) &
$\Lambda$ &
$M_{\rm crust}$ ($M_\odot$) &
$M_{\rm crust}/M$ (\%) &
$R_{\rm crust}$ (km) &
$R_{\rm crust}/R$ (\%) &
$I_{\rm crust}$ ($10^{45}\,\mathrm{g\,cm^2}$) &
$I_{\rm crust}/I$ (\%) \\
\hline

44 & 1.2 & 12.40 & 1.26 & 1220 & 0.021 & 1.78 & 1.11 & 8.95 & 0.05 & 4.12 \\
   & 1.4 & 12.49 & 1.56 &  511 & 0.019 & 1.36 & 0.93 & 7.41 & 0.05 & 3.21 \\
   & 1.6 & 12.51 & 1.88 &  225 & 0.016 & 1.03 & 0.76 & 6.08 & 0.04 & 2.13 \\
   & 2.0 & 12.36 & 2.58 &   44 & 0.012 & 0.57 & 0.51 & 4.13 & 0.04 & 1.55 \\
\hline

60 & 1.2 & 12.68 & 1.27 & 1327 & 0.021 & 1.80 & 1.14 & 8.99 & 0.06 & 4.53 \\
   & 1.4 & 12.70 & 1.60 &  536 & 0.019 & 1.36 & 0.94 & 7.40 & 0.05 & 3.34 \\
   & 1.6 & 12.69 & 1.91 &  234 & 0.016 & 0.99 & 0.77 & 6.07 & 0.05 & 2.49 \\
   & 2.0 & 12.47 & 2.59 &   44 & 0.011 & 0.55 & 0.51 & 4.09 & 0.04 & 1.35 \\
\hline

76 & 1.2 & 13.05 & 1.34 & 1508 & 0.027 & 2.23 & 1.26 & 9.68 & 0.08 & 5.78 \\
   & 1.4 & 13.01 & 1.65 &  600 & 0.023 & 1.64 & 1.02 & 7.85 & 0.07 & 4.23 \\
   & 1.6 & 12.94 & 1.96 &  254 & 0.019 & 1.21 & 0.83 & 6.44 & 0.06 & 3.09 \\
   & 2.0 & 12.61 & 2.60 &   46 & 0.013 & 0.63 & 0.55 & 4.36 & 0.04 & 1.71 \\
\hline

92 & 1.2 & 13.59 & 1.45 & 1950 & 0.029 & 2.45 & 1.35 & 9.95 & 0.09 & 6.34 \\
   & 1.4 & 13.54 & 1.76 &  776 & 0.025 & 1.76 & 1.09 & 8.02 & 0.08 & 4.61 \\
   & 1.6 & 13.45 & 2.09 &  324 & 0.021 & 1.34 & 0.90 & 6.69 & 0.07 & 3.45 \\
   & 2.0 & 13.09 & 2.75 &   59 & 0.013 & 0.68 & 0.61 & 4.47 & 0.05 & 1.87 \\
\hline

\end{tabular}%
}
\end{table*}

\begin{table*}[ht]
\centering
\caption{Global and crustal properties of neutron stars for different values 
of the symmetry energy slope ($L$) within a unified equation of state.}
\label{tab:crust_properties_unified}

\resizebox{\textwidth}{!}{%
\begin{tabular}{ccccccccccc}
\hline
$L$ (MeV) &
$M/M_\odot$ &
$R$ (km) &
$I$ ($10^{45}\,\mathrm{g\,cm^2}$) &
$\Lambda$ &
$M_{\rm crust}$ ($M_\odot$) &
$M_{\rm crust}/M$ (\%) &
$R_{\rm crust}$ (km) &
$R_{\rm crust}/R$ (\%) &
$I_{\rm crust}$ ($10^{45}\,\mathrm{g\,cm^2}$) &
$I_{\rm crust}/I$ (\%) \\
\hline

44 & 1.2 & 12.55 & 1.26 & 1221 & 0.043 & 3.60 & 1.47 & 11.71 & 0.09 & 7.04 \\
   & 1.4 & 12.60 & 1.57 &  512 & 0.038 & 2.69 & 1.20 &  9.52 & 0.08 & 5.24 \\
   & 1.6 & 12.60 & 1.90 &  226 & 0.033 & 2.05 & 1.00 &  7.97 & 0.08 & 4.08 \\
   & 2.0 & 12.42 & 2.58 &   44 & 0.023 & 1.12 & 0.67 &  5.39 & 0.06 & 2.32 \\
\hline

60 & 1.2 & 12.71 & 1.29 & 1322 & 0.049 & 4.12 & 1.44 & 11.29 & 0.11 & 8.25 \\
   & 1.4 & 12.72 & 1.60 &  539 & 0.043 & 3.08 & 1.18 &  9.28 & 0.10 & 6.19 \\
   & 1.6 & 12.70 & 1.92 &  234 & 0.037 & 2.32 & 0.97 &  7.65 & 0.09 & 4.67 \\
   & 2.0 & 12.47 & 2.58 &   44 & 0.025 & 1.24 & 0.64 &  5.13 & 0.07 & 2.59 \\
\hline

76 & 1.2 & 12.95 & 1.33 & 1502 & 0.034 & 2.81 & 1.23 & 9.50 & 0.07 & 5.62 \\
   & 1.4 & 12.94 & 1.64 &  601 & 0.028 & 2.03 & 1.00 & 7.73 & 0.07 & 4.26 \\
   & 1.6 & 12.88 & 1.96 &  254 & 0.024 & 1.50 & 0.81 & 6.29 & 0.06 & 3.09 \\
   & 2.0 & 12.58 & 2.59 &   46 & 0.016 & 0.80 & 0.54 & 4.29 & 0.04 & 1.70 \\
\hline

92 & 1.2 & 13.42 & 1.44 & 1955 & 0.022 & 1.82 & 1.12 & 8.30 & 0.05 & 3.63 \\
   & 1.4 & 13.41 & 1.75 &  778 & 0.019 & 1.31 & 0.91 & 6.79 & 0.05 & 2.76 \\
   & 1.6 & 13.35 & 2.08 &  327 & 0.016 & 0.99 & 0.75 & 5.62 & 0.04 & 2.02 \\
   & 2.0 & 13.02 & 2.74 &   59 & 0.011 & 0.52 & 0.48 & 3.67 & 0.03 & 1.10 \\
\hline

\end{tabular}%
}
\end{table*}

Regarding the size of the crust, $R_{\rm crust}$, there are both similarities and differences. The direct connection predicts a smaller contribution to the radius, as in the case of the mass, but now the results do not decrease monotonically with the slope. There is a small increase in $R_{\rm crust}$ at $L = 92$ MeV. On the other hand, there is no peak in $R_{\rm crust}$ within the unified scheme, and the size of the crust always decreases with $L$, while the smooth crossover shows the opposite behaviour, with $R_{\rm crust}$ increasing monotonically with $L$.

Quantitatively, the contribution of the crust to the total radius of the star is significantly larger than the contribution to the mass. The crust size can reach over 8$\%$ already in the direct-connection, and can surpass 11\% for the unified scheme.

The last crustal quantity analyzed here is the relative contribution of the crust to the MOI. The results are displayed in Fig.~\ref{relativeMOI}. Since $I \propto MR^2$, the influence of the crust in the MOI is unsurprisingly intermediate between the influence of the crust in the mass and in the radius of the NS. Here again, the direct connection produces the lower contribution for the crust compared with the two other approaches. Furthermore, the contribution of the crust to the MOI decreases with the slope and with the star mass, as happened with  $M_{\rm crust}$. The contribution for the direct connection is essentially small, barely reaching 1\%.

In the case of smooth crossover, we notice a significant increase of the crustal MOI, now reaching values as high as 6\%. The dependence on the slope is once more reversed when compared with direct connection. 

Finally, we have a more complex behaviour within the unified EOS, once there is a peak in $I_{\rm crust}$ for $L = 60$ MeV, which was also present in the analyzes of $M_{\rm crust}$. Compared with the smooth crossover, we see that the crossover approach underestimates the crust MOI for $L = 44$ and 60 MeV; it reproduces the same contribution for $L = 76$  MeV, and overestimates $I_{\rm crust}$ for $L = 92$ MeV.

We conclude this section by noting that although different approaches cause only small to moderate changes in the global properties of the NS, they have an important effect to the crustal properties. It is also worth mentioning that in the absence of a unified EOS, the direct connection can be safely used in the mass-radius diagram, as well as in the calculation of the MOI. However, the smooth crossover is imperative in the calculation of the dimensionless tidal parameter, $\Lambda$, corroborating the discussion presented in Ref.~\cite{Pang2026PRD}. Smooth crossover is also a better option for the crustal properties.

\newpage

\section{Conclusions}
\label{sec:conclusions}

We have compared three treatments of the NS crust (a direct connection, a smooth crossover, and a unified EOS) for four values of the symmetry energy slope, $44 \le L \le 92$ MeV. 
Our main findings are:
\begin{itemize}
\item \textit{Neutron drip and crust--core transition.} In the unified EOS, the neutron-drip density increases with $L$ (from $2.49$ to $3.42\times10^{-4}$ fm$^{-3}$), the crust--core transition density present a reverse behaviour  and decreases (from $0.088$ to $0.058$ fm$^{-3}$), while the transition pressure is non-monotonic, peaking at $L = 60$ MeV.

\item \textit{Global properties.} The differences among crust treatments decrease with stellar mass: the radius agrees within 170 m for a $1.2\,M_\odot$ star, 130 m for a $1.4\,M_\odot$ star, and 70 m for a $2.0\,M_\odot$ star, and the moment of inertia within 2\%, 1\%, and less than 1\%, respectively. 
These differences remain well below those arising from the variation of $L$, which changes the radius of a $1.4\,M_\odot$ star by 0.8--0.9 km. The exception is the tidal deformability, which the direct connection underestimates by about 21\% for $L = 92$ MeV at both $1.2$ and $1.4\,M_\odot$. An unified or a smooth crossover EOS is therefore mandatory
in the analyses of the dimensionless tidal parameter.

\item \textit{Crustal properties.} Relative to the unified EOS, the direct
connection yields a crust that is 15--36\% thinner, 6--12 times less
massive, and that carries a fraction of the total moment of inertia 5--8
times smaller. These differences are comparable to or larger than those
arising from the variation of $L$, largely because the direct connection
ends the crust at $0.01$ fm$^{-3}$, well below the transition densities of
the unified EOS.
\item \textit{Smooth crossover.} It reproduces the global properties of the
unified EOS, with tidal deformabilities within 0.5\%, but not the crustal
ones: the crustal fraction of the moment of inertia is underestimated by
about 40\% for $L = 44$ MeV and overestimated by about 70\% for
$L = 92$ MeV.
\item \textit{Dependence on $L$.} The trend of the crustal properties with
$L$ is set by the crust treatment. The crustal fractions decrease
monotonically with $L$ for the direct connection and increase for the
crossover. For the unified EOS they are non-monotonic and follow the
crust--core transition pressure. Correlations between $L$ and crustal
properties inferred from non-unified EOSs should therefore be treated with
caution.
\end{itemize}

The crustal fractions of the moment of inertia obtained with the direct
connection and with the unified EOS lie on opposite sides of the value of
about 1.4\% often quoted as necessary for a crustal origin of pulsar
glitches~\cite{Link1999}. Since our crust model includes neither neutron
superfluidity nor entrainment, this is not a statement about glitches
themselves. It does illustrate, however, that the crust treatment alone can
change the conclusions drawn from crust-based astrophysical constraints,
which should therefore rely on unified EOSs.
Finally, we will make the four unified EoSs produced in this work available to the community in the CompOSE database~\cite{CompOSECoreTeam:2022ddl}.

\vspace{1cm}

\appendix

\textbf{Acknowledgments:}L.L.L.  was partially supported by CNPq (Brazil)
under Grant No 305347/2024-1, and INCT-FNA Project 408419/2024-5

\newpage

$$ $$

%%%%%%%%%%%%%%%%%%%%%%%%%
\bibliography{aref}
%%%%%%%%%%%%%%%%%%%%%%%%%%%%%

\end{document}